\documentclass[aps,pra,amsmath,amssymb,floatfix,twocolumn,amsmath,superscriptaddress,twocolumn,nofootinbib,tighten,letterpaper]{revtex4}
\usepackage{multirow}
\usepackage{bbold}
\usepackage{subfigure}
\usepackage{color}
\usepackage{mathrsfs}
\usepackage{hyperref}
\usepackage[normalem]{ulem}
\usepackage{bm}
\usepackage{comment}

\renewcommand\vec[1]{\ensuremath\mathrm{\mathbf{#1}}}

\usepackage{amsfonts, relsize, color}
\usepackage{graphics}
\usepackage{graphicx}
\usepackage{subfigure}
\usepackage{hyperref}
\usepackage{color}

\renewcommand{\vec}{\mathbf}
\newcommand{\D}{\mathrm{d}}

\graphicspath{{FinalFigures/}}

\usepackage{xr}
\begin{document}
\title{Dirty higher-order Dirac semimetal: Quantum criticality and bulk-boundary correspondence}

\author{Andr\' as L. Szab\' o}
\affiliation{Max-Planck-Institut f\"{u}r Physik komplexer Systeme, N\"{o}thnitzer Str. 38, 01187 Dresden, Germany}

\author{Bitan Roy}
\affiliation{Department of Physics, Lehigh University, Bethlehem, Pennsylvania, 18015, USA}

\date{\today}

\begin{abstract}
We analyze the stability of time-reversal (${\mathcal T}$) and lattice four-fold ($C_4$) symmetry breaking three-dimensional higher-order topological (HOT) Dirac semimetals (DSMs) and the associated one-dimensional hinge modes in the presence of random pointlike charge impurities. Complementary real space numerical and momentum space renormalization group (RG) analyses suggest that a HOTDSM, while being a stable phase of matter for sufficiently weak disorder, undergoes a continuous quantum phase transition into a trivial metal at finite disorder. However, the corresponding critical exponents (numerically obtained from the scaling of the density of states) are extremely close to the ones found in a dirty, but first-order DSM that on the other hand preserves ${\mathcal T}$ and $C_4$ symmetries, and support two Fermi arc surface states. This observation suggests an emergent \emph{superuniversality} (insensitive to symmetries) in the entire family of dirty DSMs, as also predicted by a leading-order RG analysis. As a direct consequence of the bulk-boundary correspondence, the hinge modes in a system with open boundaries gradually fade away with increasing randomness, and completely dissolve in the trivial metallic phase at strong disorder.      
\end{abstract}

\maketitle

\section{Introduction}

Traditionally, the bulk-boundary correspondence in a $d$-dimensional topological system refers to the boundary modes residing on a $(d-1)$-dimensional interface, also characterized by the codimension $d_c=d-(d-1)=1$~\cite{kane-hasan-review, Qi-zhang-review, bernevig-hughes-book, chingkai-review, armitage-review}. This notion has recently been generalized to encompass boundary modes with $d_c >1$, e.g., pointlike corner ($d_c=d$) and one-dimensional hinge ($d_c=d-1$) modes, which led to the construction of insulating (electric and thermal) and nodal higher-order topological (HOT) phases~\cite{benalcazar2017, schindler2018, song2017, benalcazar-prb2017,langbehn2017,franca2018,schindler-sciadv2018,wang-PRL2019,hsu2018,wang1-2018,hughes-HOTDSM,calugaru2019,ghorashi2019,Klinovaja2019,Tnag2019,Klinovaja2019PRR,ZYan2019,roy-singleauthor2019,bernevig-NatComm2020,VLiu-PRL-2020,andras-2019,dassarma-HOTSC}. A question regarding the stability of such exotic phases of matter in the presence of interactions and/or disorder arises naturally. Due to a finite gap in the quasiparticle spectra, while the HOT insulators are robust against sufficiently weak interactions and disorder, their influences on gapless HOT phases demand more careful analyses.

\begin{figure}[t!]
\includegraphics[width=0.47\linewidth]{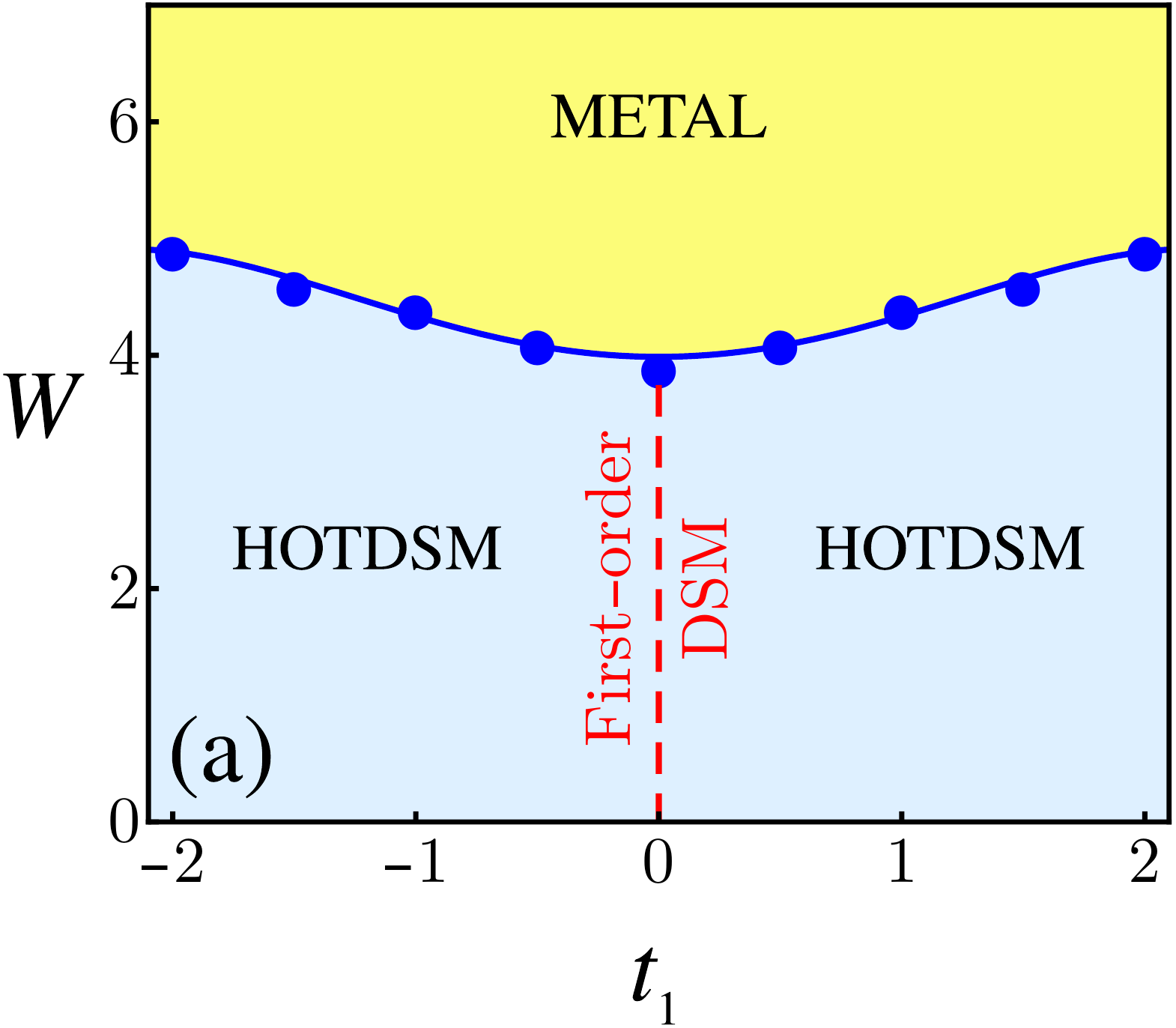}
\includegraphics[width=0.475\linewidth]{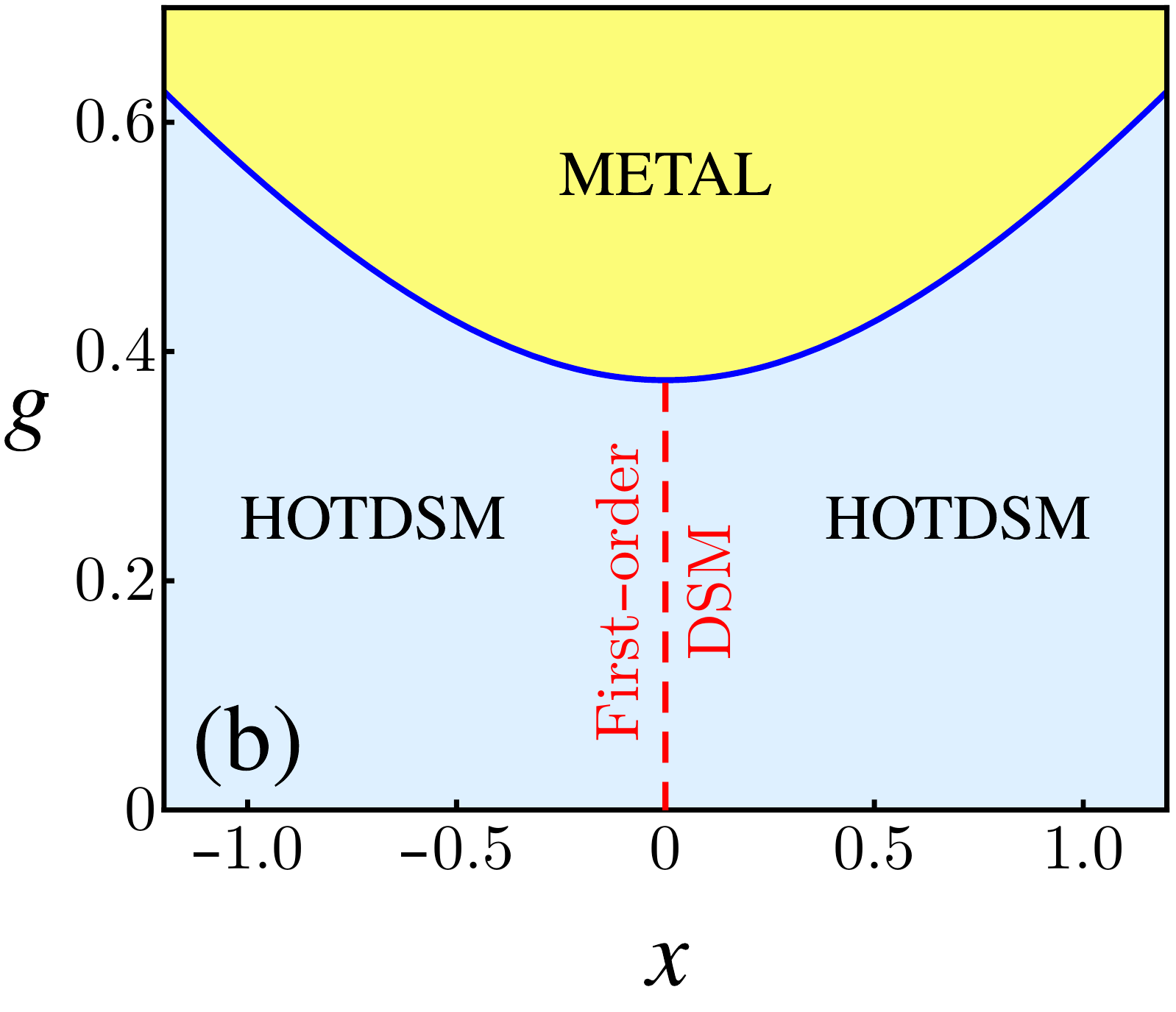}
\caption{Phase diagrams of a dirty HOTDSM obtained from (a) the scaling of DOS at zero energy (numerically) in a lattice model [Eq.~(\ref{Eq:tb})], and (b) a leading-order RG analysis of the continuum model [Eq.~(\ref{Eq:hDP})], which are in qualitative agreement. A first-order DSM is realized along the red dashed lines (a) $t_1=0$ or (b) $x=0$, where $x=b \Lambda/v$, with $\Lambda (v)$ as the ultraviolet cutoff (Fermi velocity), and $v \sim t a$, $b \sim t_1 a^2$, $\Lambda \sim a^{-1}$ ($a$ is the lattice spacing). When disorder $W$ or $g$ (dimensionless disorder coupling in the continuum limit) is weak HOT and first-order DSMs are stable. But, at stronger disorder they undergo a QPT into a metallic phase, where DOS at zero energy is finite. The blue dots in (a) are numerically obtained transition points between DSM and metal, across which we find single-parameter scaling of DOS [Fig.~\ref{Fig:Collapse}].
}~\label{Fig:fig1}
\end{figure}

\begin{figure*}[t!]
\centering
\includegraphics[width=0.70\linewidth]{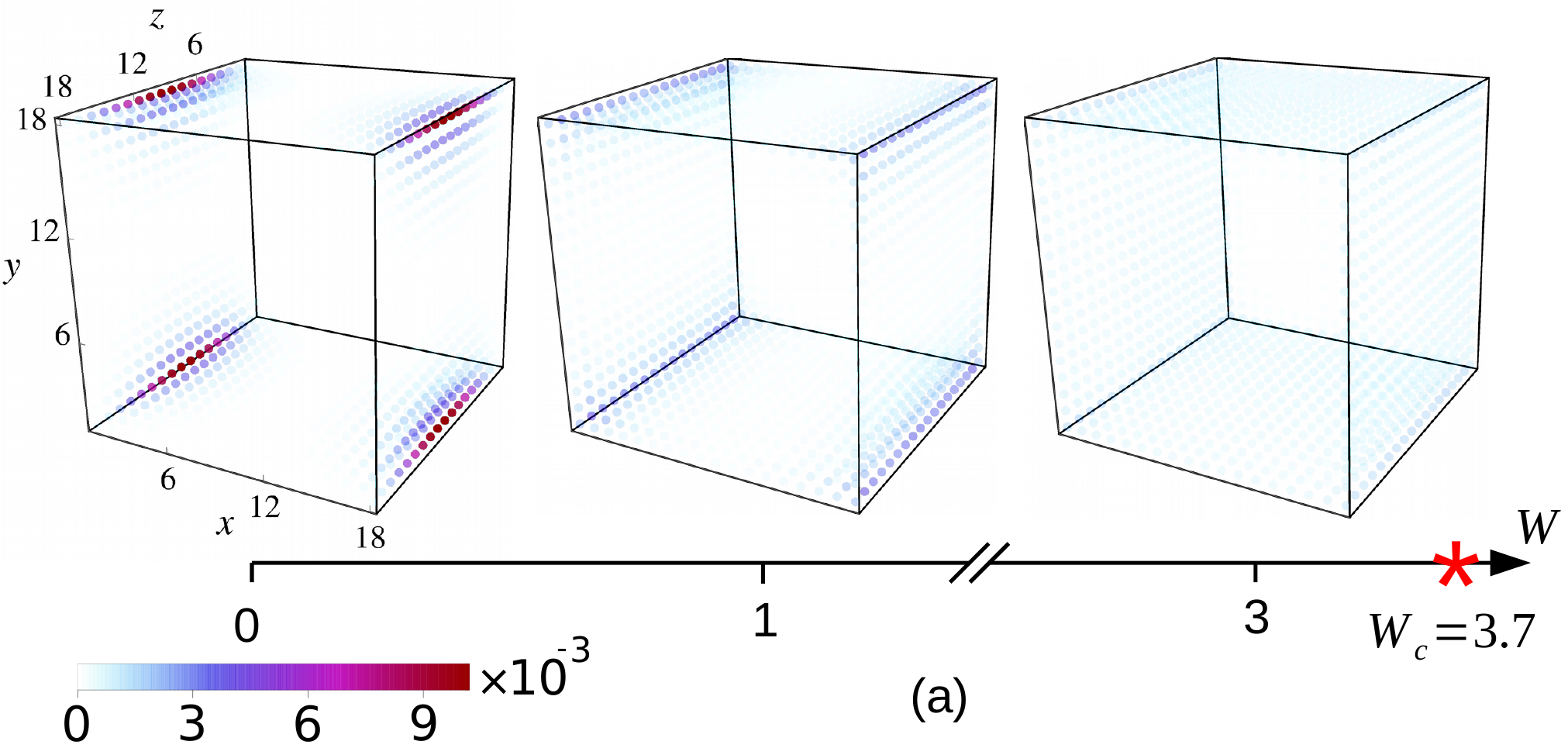}
\includegraphics[width=0.25\linewidth]{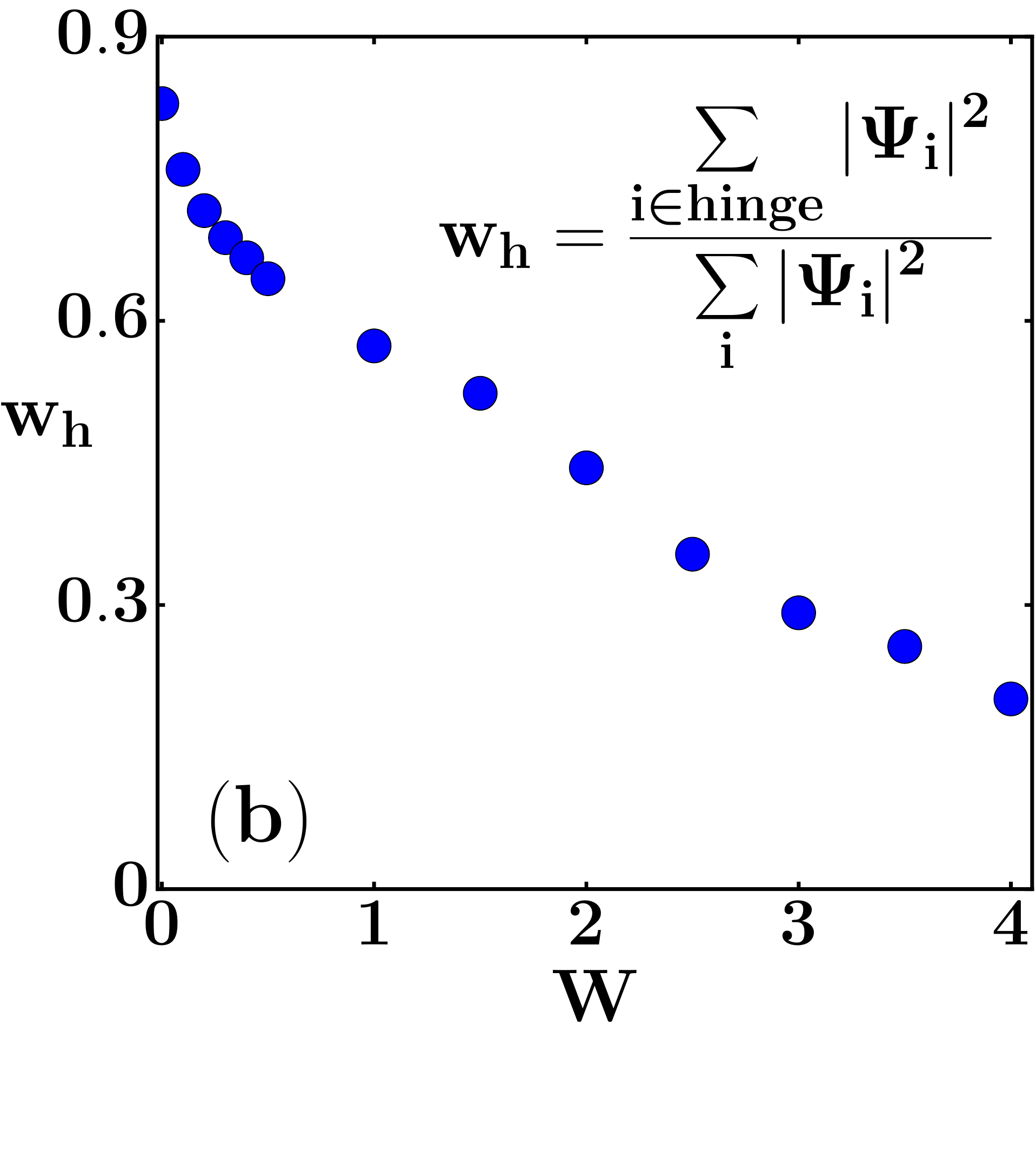}
\caption{(a) Melting of topological hinge states of a HOTDSM with increasing disorder ($W$) in an open cubic system with $L=18$, obtained by averaging the local DOS over $300$ independent disorder realizations. To recover the particle-hole symmetry (on average), we switch off the $\Gamma_5$ term in the lattice model [see Eq. (\ref{Eq:tb})], which is not responsible for the hinge states when the crystal is cleaved such that its four corners are at $(\pm \frac{L}{2},\pm \frac{L}{2})$ for any $z$. The critical disorder $W_c=3.7 \pm 0.1$ is obtained from the scaling of average DOS in a periodic system of $L=200$. The dissolution of the hinges into the bulk sets in for $W<W_c$ due to the finite size effects. (b) Scaling of the hinge localization ($W_h$) of the topological hinge modes with increasing disorder ($W$), confirming its gradual melting.
}~\label{Fig:FadingHinge}
\end{figure*}

Here we focus on a three-dimensional Higher-order topological Dirac semimetal (HOTDSM), supporting one-dimensional hinge modes in open systems, and scrutinize its stability when littered with pointlike quenched random charge impurities. From complementary real space numerical analyses in periodic systems and field-theoretic momentum space renormalization group (RG) calculation, we conclude that HOTDSM is a stable phase of matter in the presence of sufficiently weak disorder. But, the system undergoes a continuous quantum phase transition (QPT) into a trivial diffusive metal at finite disorder, where the density of states (DOS) at zero energy is finite, while it vanishes in HOTDSM. We arrive at these conclusions from the scaling of the average DOS, computed using the kernel polynomial method (KPM)~\cite{KPM-RMP}. The phase diagrams obtained from these two complementary methods are shown in Figs.~\ref{Fig:fig1}(a) and~\ref{Fig:fig1}(b), respectively, which are in qualitative agreement. As a direct consequence of the bulk-boundary correspondence, the associated topological hinge modes in open systems gradually fade away with increasing randomness and completely melt into a trivial metallic bulk for sufficiently strong disorder, as shown in Fig.~\ref{Fig:FadingHinge}. Therefore, sharp hinge localized topological modes can be observed at least in weakly disordered HOTDSMs, via angle-resolved photoemission spectroscopy (ARPES) and scanning tunneling microscope (STM).  

Moreover, numerically extracted values for the dynamical scaling exponent (DSE) and correlation length exponent (CLE) across a broad range of parameters (that also includes first-order DSMs) appear to be \emph{almost} the same (within the numerical accuracy) [see Table \ref{tab:exponents}], and yield satisfactory single-parameter data collapses across the (HOT)DSM-metal continuous QPT [see Fig.~\ref{Fig:Collapse}]. To this end we note that while a HOTDSM breaks time-reversal (${\mathcal T}$) and lattice four-fold ($C_4$) rotational symmetries, a first-order DSM preserves them. Therefore, this observation in turn strongly promotes the notion of an emergent \emph{superuniversality} near a diffusive quantum critical point (QCP) in the entire family of dirty DSMs, irrespective of their symmetries and the topological order. We also substantiate these findings from a leading-order momentum space RG analysis.

The rest of the paper is organized as follows. In the next section, we introduce the lattice model for an HOTDSM and discuss its symmetries. In Sec.~\ref{sec:numerics}, we numerically analyze the scaling of DOS. Sec.~\ref{sec:RG} is devoted to the RG analysis of disordered HOTDSM. In Sec.~\ref{sec:summary}, we summarize the results and discuss related problems. Additional technical details are relegated to the appendices, and additional numerical data are shown in the Supplementary Materials~\cite{SM}.

\section{Lattice model and symmetries}~\label{sec:latticemodel}

We set out by considering a tight-binding model for a three-dimensional HOTDSM that in the momentum space reads $\hat{h}^\vec{k}=\hat{h}^{\bf k}_0+\hat{h}^{\bf k}_1$, where~\cite{hughes-HOTDSM,calugaru2019, bernevig-NatComm2020, andras-2019}
\allowdisplaybreaks[4]
\begin{eqnarray}~\label{Eq:tb}
\hat{h}^{\bf k}_0 &=& t \sum_{j=1}^2 S_j \Gamma_j + \big[ t_z C_3 - m_z + t_0\sum_{j=1}^2 (1-C_j) \big] \Gamma_3, \nonumber \\
\hat{h}^{\bf k}_1 &=& t_1 \big[ (C_1-C_2)\Gamma_4 + S_1S_2 \Gamma_5 \big],
\end{eqnarray}
$C_j\equiv\cos(k_j a)$, $S_j\equiv\sin(k_j a)$, $k_j$ being the components of momenta and $a$ is the lattice spacing that we set to unity. Here, $i=1,2,3$ correspond to $x,y,z$ respectively. The four-component Hermitian $\Gamma$ matrices, $\Gamma_1=\sigma_3 \tau_1$, $\Gamma_2=\sigma_0 \tau_2$, $\Gamma_3=\sigma_0 \tau_3$, $\Gamma_4=\sigma_1 \tau_1$, $\Gamma_5=\sigma_2 \tau_1$, satisfy the anticommuting Clifford algebra $\{\Gamma_i,\Gamma_j\}=2 \delta_{ij}$, for $i,j=1, \cdots, 5$. Pauli matrices $ \sigma_\mu$ ($\tau_\mu$) operate on the spin (orbital) degrees of freedom, where $\mu=0, \cdots, 3$. For the rest of this paper we set $m_z=0$ and $t=t_0=t_z=1$. Then for $t_1=0$, the above model can be viewed as stacked two-dimensional quantum spin Hall insulators (QSHIs) along the $k_z$ axis, and each layer accommodates two counter-propagating one-dimensional edge states for opposite spin projections. The resulting three-dimensional system supports two Dirac-points at $\vec{k}=(0,0,\pm \frac{\pi}{2}) \equiv \pm {\bf K}$, and ribbon like edge modes localized on the $xz$ and $yz$ planes, yielding two copies of Fermi arcs connecting the Dirac points on the $k_xk_z$ or $k_yk_z$ plane. This system describes a first-order DSM, leading to a $\rho(E)\sim |E|^2$ scaling of DOS at low energies. A first-order DSM preserves the time-reversal symmetry: ${\mathcal T}=\sigma_2 \tau_0 {\mathcal K}$, where ${\mathcal K}$ is the complex conjugation. The first-order DSM also preserves the parity (${\mathcal P}$) symmetry, under which ${\bf r} \to -{\bf r}$ and ${\mathcal P}=\sigma_0 \tau_3$.

\begin{figure*}[t!]
\centering
\includegraphics[height=3.0cm]{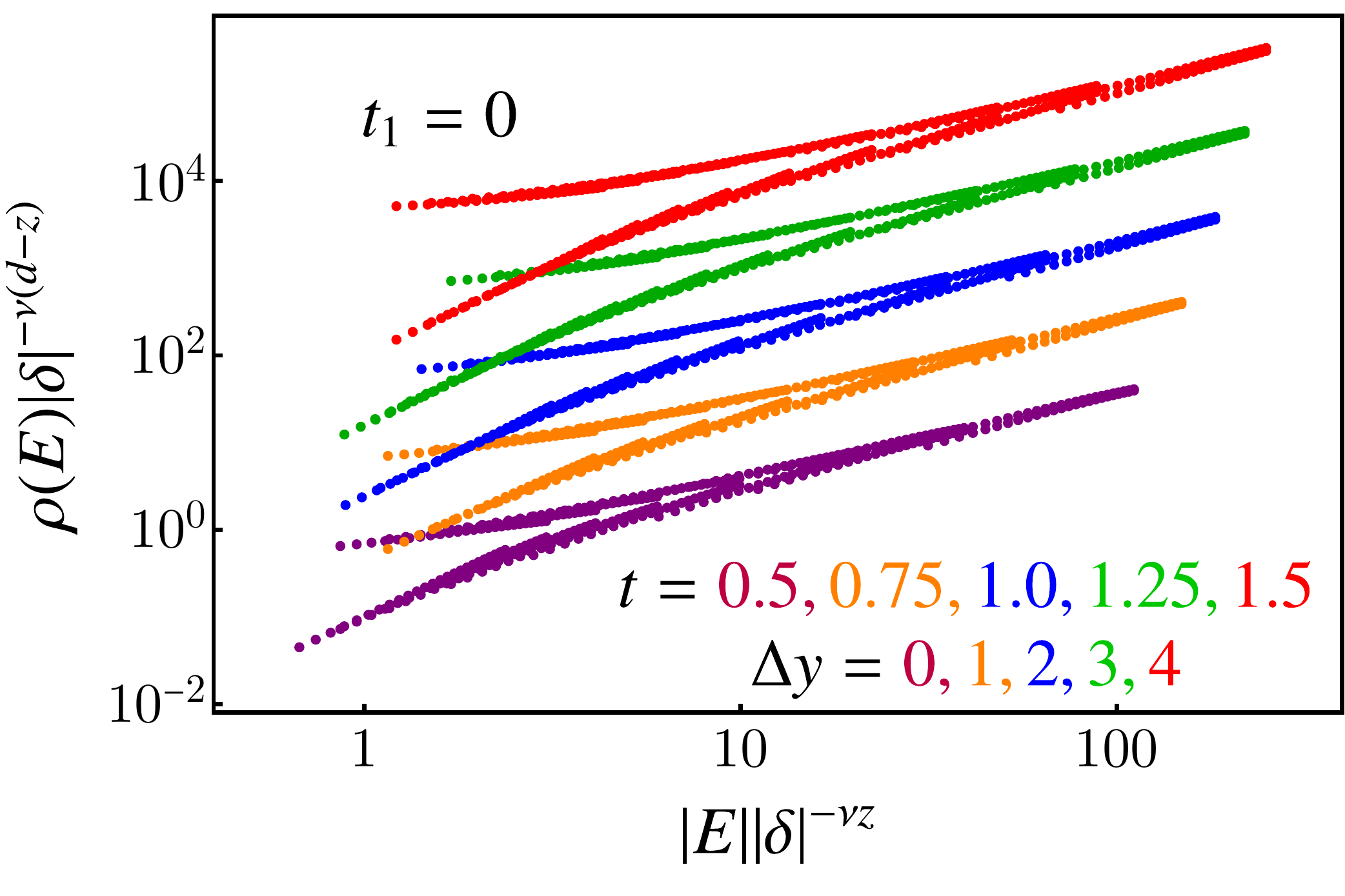}
\includegraphics[height=3.0cm]{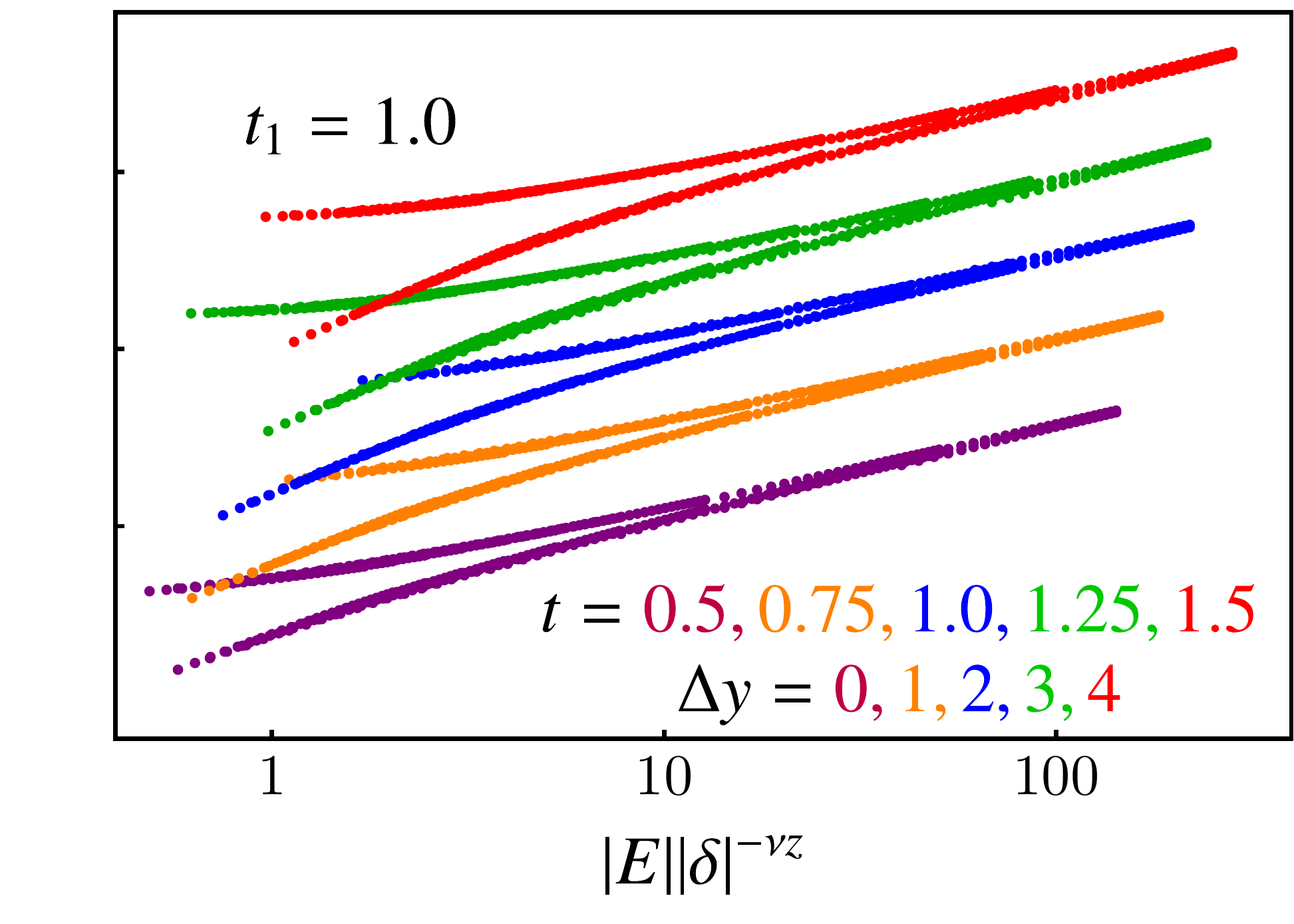}
\includegraphics[height=3.0cm]{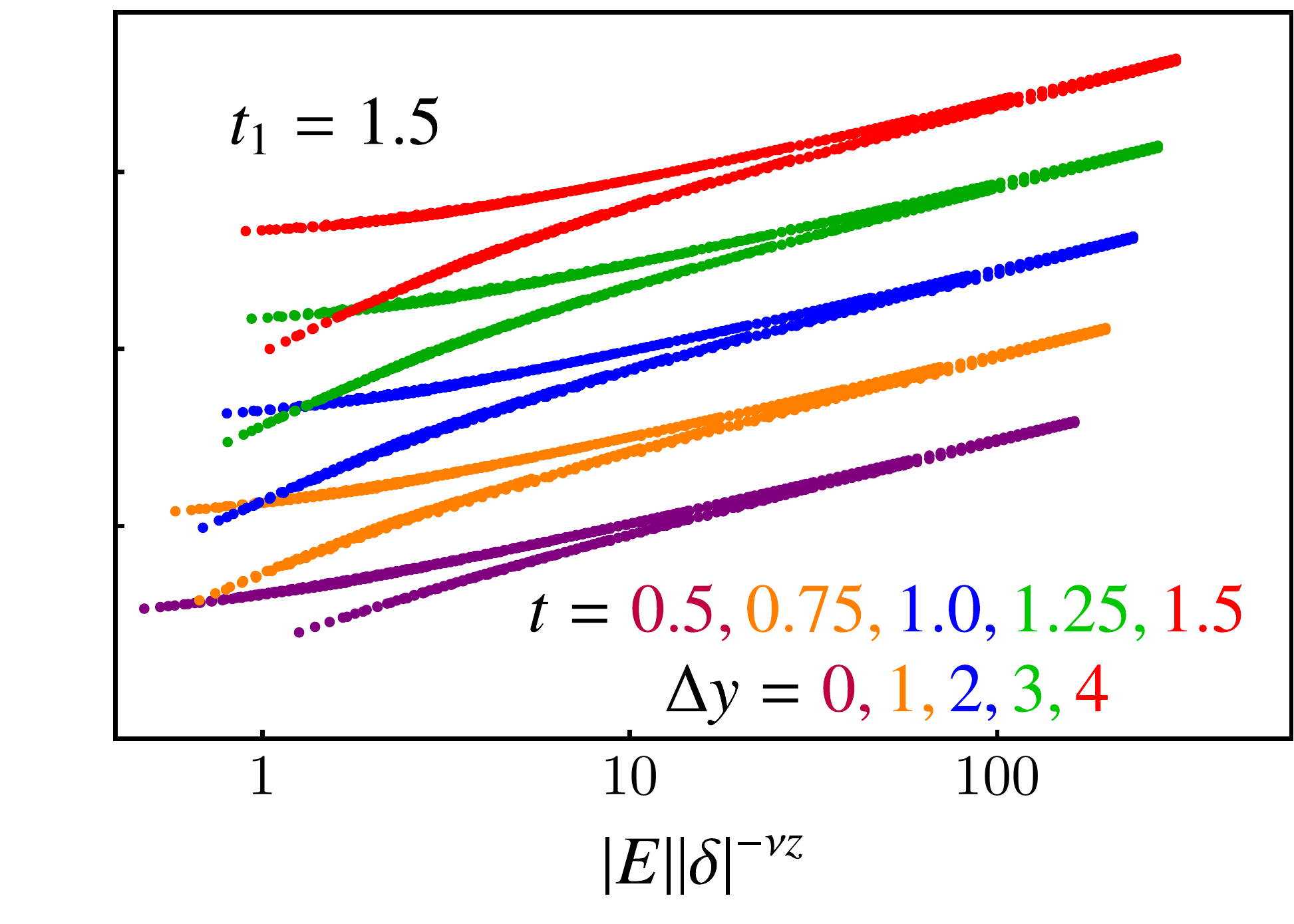}
\includegraphics[height=3.0cm]{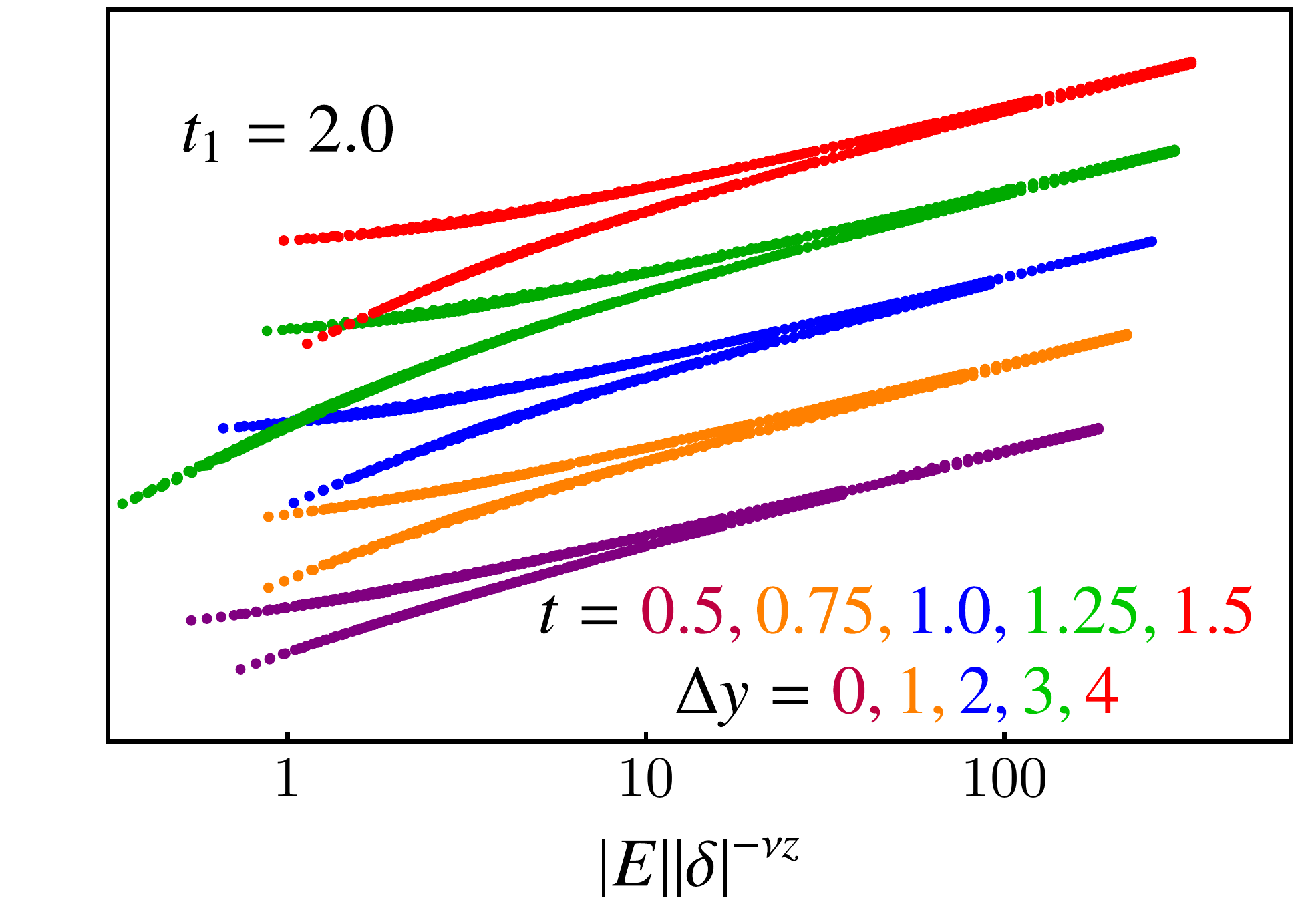}\\
\includegraphics[height=3.0cm]{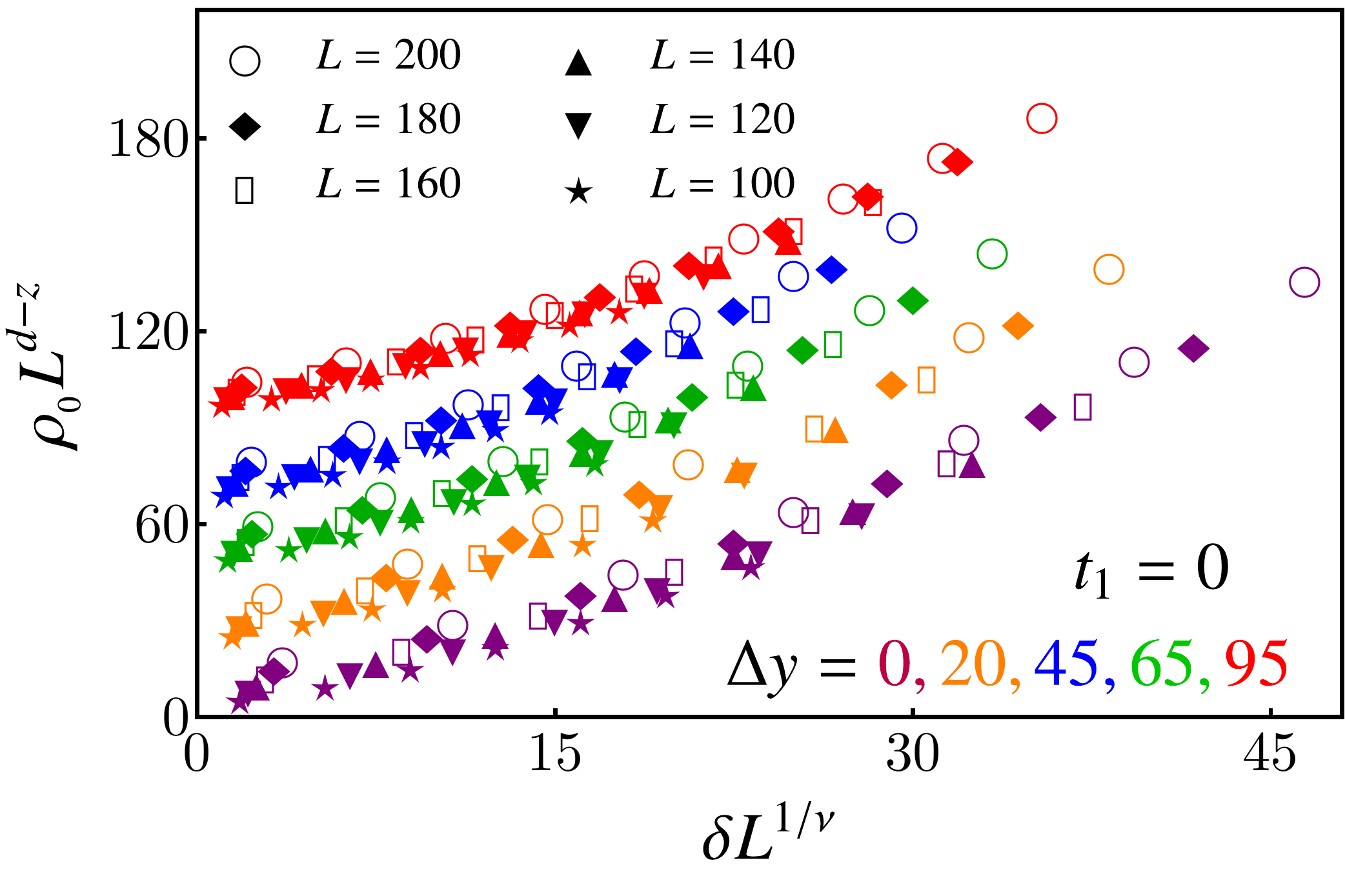}
\includegraphics[height=3.0cm]{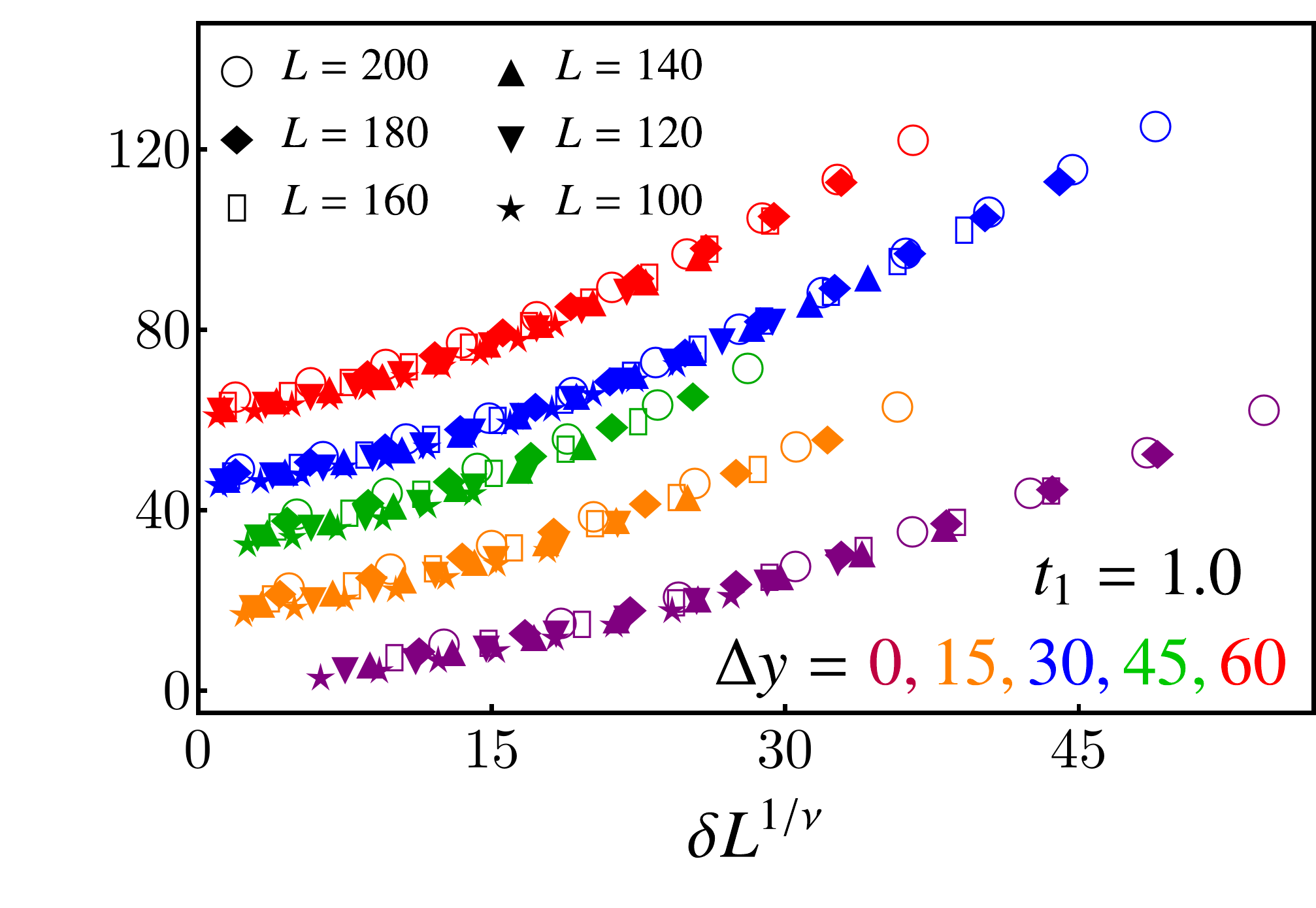}
\includegraphics[height=3.0cm]{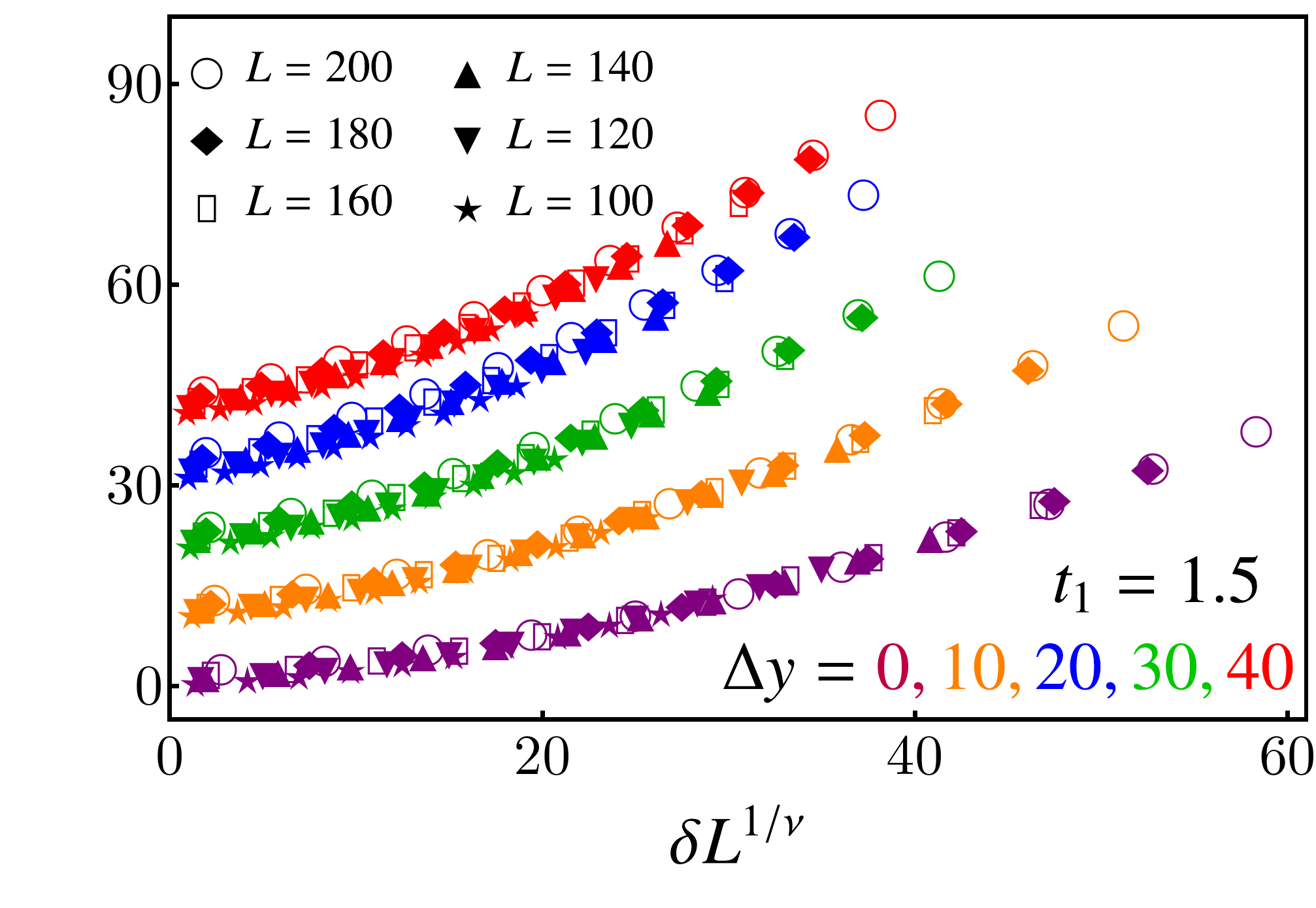}
\includegraphics[height=3.0cm]{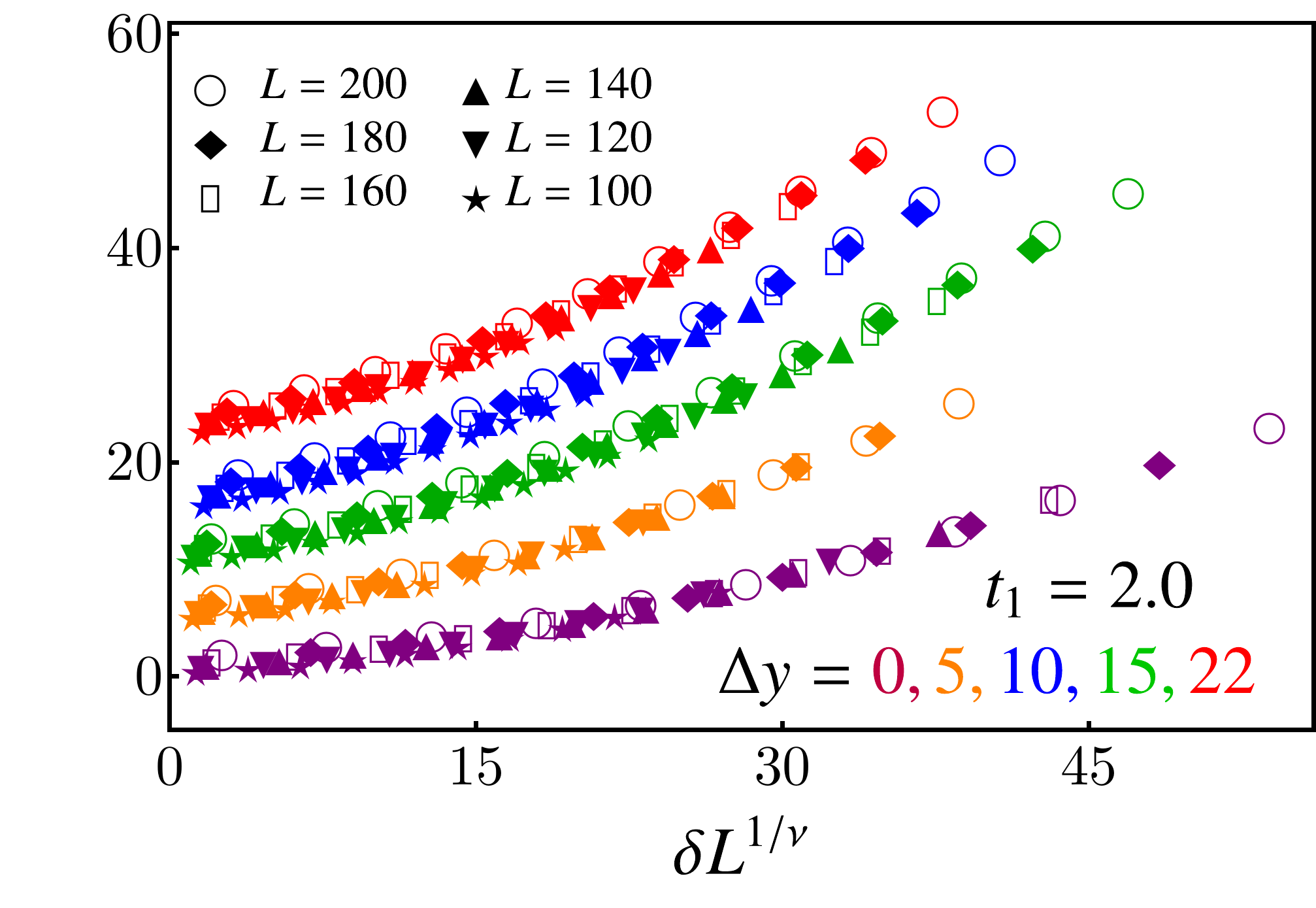}
\caption{Top: Data collapses for $\rho(E)$ in a system of $L=200$. For clarity, we shift the data sets for different values of $t$ according to $\log_{10} [\rho(E)|\delta|^{-\nu(d-z)}]+\Delta y$ (quoted in each panel). All data sets collapse on three branches. The semimetallic (lower) and metallic (upper-left) branches meet inside the quantum critical regime (upper-right).
Bottom: Finite size data collapses for $\rho(0) (\equiv \rho_0)$ inside the metallic phase. We vertically shift the data sets for different values of $t$ according to $\rho_0L^{d-z}+\Delta y$. 
}~\label{Fig:Collapse}
\end{figure*}

On the other hand, $\hat{h}^{\bf k}_1$ breaks the time-reversal symmetry and acts as a lattice $C_4$ symmetry breaking, momentum-dependent or Wilson mass in each layer of QSHI. It changes sign under the $C_4$ rotations and thus acts as a domain wall mass; resulting in four corner-localized zero-energy states (with $d_c=2$) in each two-dimensional insulating layer, according to a generalized Jackiw-Rebbi index theorem~\cite{jackiw-rebbi}. Stacking such layers of two-dimensional HOT insulators in the momentum space along the $k_z$ axis gives rise to the one-dimensional hinge modes along the $z$ direction~\cite{andras-2019} [Fig.~\ref{Fig:FadingHinge}]. We then realize a second-order DSM, as $\hat{h}^{\bf k}_1$ does not affect the Dirac nodes at $\pm {\bf K}$, where it vanishes.

The HOTDSM also breaks the parity (${\mathcal P}$) symmetry. We then define a pseudo time-reversal symmetry ${\mathcal T}^\prime= {\mathcal P} {\mathcal T}$, such that $({\mathcal T}^\prime)^2=-1$. The HOTDSM preserves such pseudo time-reversal symmetry, which in turn assures the Kramers degeneracy of the valence and conduction bands~\cite{roy-singleauthor2019}.

As $\hat{h}^{\bf k}_1$ vanishes \emph{quadratically} with momentum around the Dirac points, it only reduces the DOS at sufficiently low energies without altering its overall $\rho(E) \sim |E|^2$ scaling [Fig.~\ref{Fig:fig4}(a)]. The subdominant influence of the Wilson mass on DOS suggests that the HOTDSM is stable for sufficiently weak disorder, and enters into a metallic phase via a QPT at finite disorder, \emph{qualitatively} similar to the situation in first-order Dirac and Weyl semimetals~\cite{fradkin,goswami-chakravarty-1, ominato-kishino, roy-dassarma-1, syzranov-1, syzranov-2, roy-dassarma-2, goswami-chakravarty-2, carpentier-1, carpentier-2, herbut, brouwer-1, brouwer-2, pixley-1,pixley-2,ohtsuki-shindou,bera-sau-roy, larsfritz,roy-slager-juricic-1,carpetier-3, mirlin, carpentier-4,ogata}, up to exponentially small, but debated rare region effect~\cite{nandkishore-RR, huses-pixley-RR, altland-2}. Next we anchor these anticipations by implementing the tight-binding model from Eq.~(\ref{Eq:tb}) on a cubic lattice with periodic boundary in each direction and numerically computing the DOS using the KPM, and investigate the critical properties of the HOTDSM-metal QPT. We consider pointlike charge impurities, which is the dominant source of elastic scattering in any real material, distributed uniformly and independently within the range $[-W/2,W/2]$ at each site of the cubic lattice. Numerically obtained critical exponents are summarized in Table~\ref{tab:exponents}, which we use to perform a finite energy (size) data collapse, shown in Fig.~\ref{Fig:Collapse} top (bottom).


\section{Numerical analysis of Density of states}~\label{sec:numerics}

To formulate the scaling theory for DOS, we concentrate around the dirty QCP at $W=W_c$, and parametrize the distance from it by $\delta=(W-W_c)/W_c$. The number of states $N(E,L)$ in a $d$-dimensional system of linear size $L$ below some energy $E$ is in general a function of two dimensionless parameters, $L/\xi$ and $E/E_0$. Here $\xi$ is the correlation length, which diverges at the QCP as $\xi\sim \delta^{-\nu}$, and $E_0 \sim\xi^{-z}$ is the corresponding energy scale. Since the number of states is proportional to $L^d$, the functional form of $N(E,L)$ ought to be~\cite{herbut, pixley-2, bera-sau-roy}
\begin{equation}
N(E,L)=(L/\xi)^d \; F(E\xi^z, L/\xi),
\end{equation}
where $F(x,y)$ is an unknown, but universal function of its arguments. The DOS is then given by 
\begin{equation}~\label{Eq:Scaling_DOSgeneral}
\rho(E)=\frac{1}{L^d}\frac{\D N(E,L)}{\D E}=\delta^{(d-z)\nu} \; G(|E|\delta^{-z\nu}, L^{1/\nu}\delta).
\end{equation}
To investigate the scaling behavior of the universal function $G$, we consider the scaling of $\rho(E)$ at low energies inside the HOTDSM and metallic phases, as well as inside the critical regime around the QCP at $W=W_c$. For now we assume $L$ to be sufficiently large, such that the $L$-dependence of $G$ can be neglected.

\begin{figure}[t!]
\includegraphics[width=0.95\linewidth]{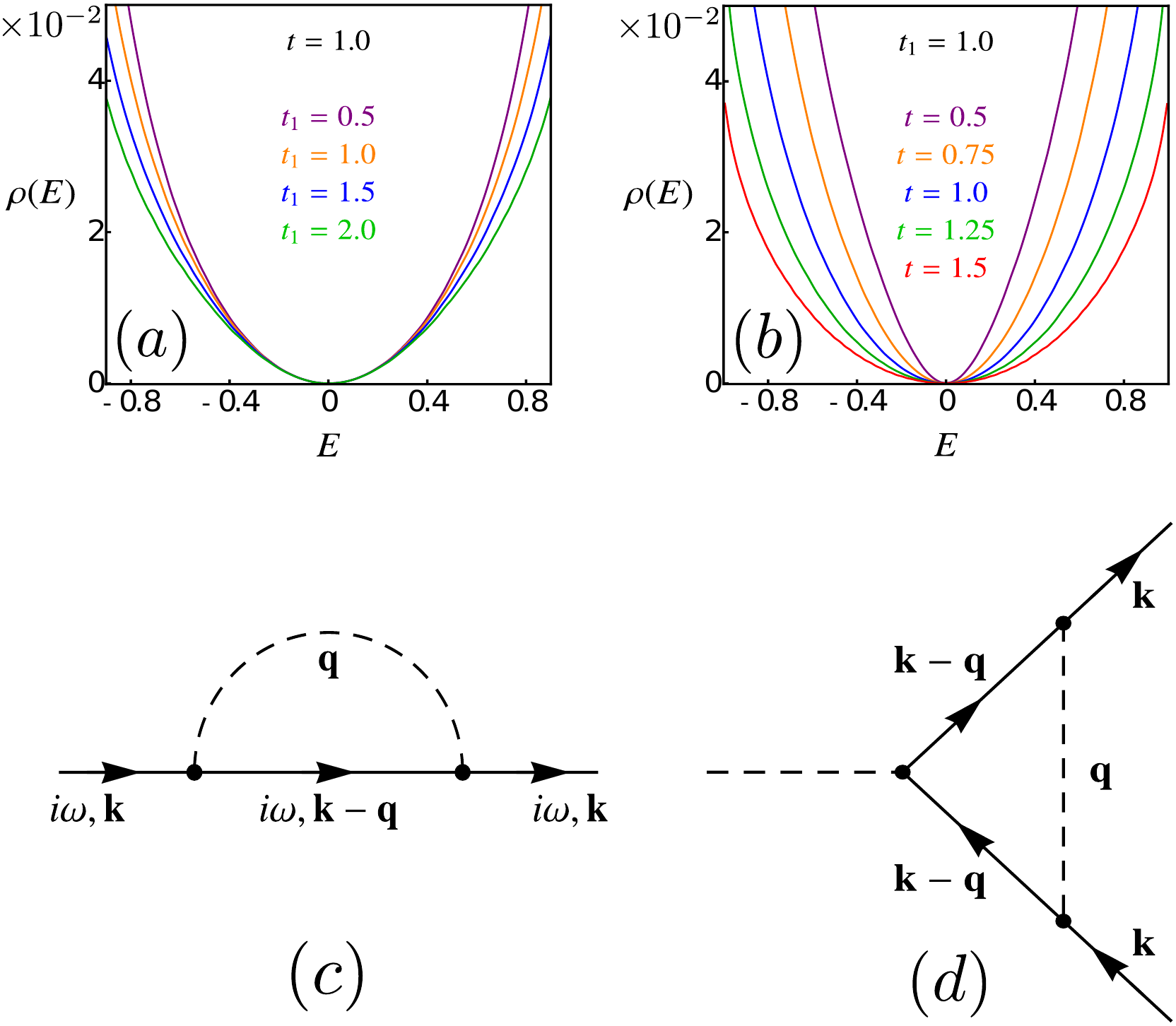}
\caption{Scaling of DOS $\rho(E)$ with (a) the $C_4$ symmetry breaking Wilson mass ($t_1$), yielding a HOTDSM, for fixed $t$ or Fermi velocity and (b) $t$ for fixed $t_1$ in a cubic system of $L=200$ with periodic boundaries for $W=0.05$. With increasing $t_1$ or $t$ the DOS decreases, without altering the $\rho(E) \sim |E|^2$ scaling, leading to the enhancement of critical disorder strength ($W_c$) for metallicity [Fig.~\ref{Fig:fig1} and Table~\ref{tab:exponents}]. Feynman diagrams contributing to the leading-order RG analysis are shown in (c) and (d). Solid (dashed) lines represent fermion (disorder) fields.   
}~\label{Fig:fig4}
\end{figure}

For linearly dispersing HOT Dirac fermions the DOS at low energies scales as $\rho(E)\sim |E|^{d-1}$ [see Figs.~\ref{Fig:fig4}(a) and ~\ref{Fig:fig4}(b)] when $\delta<0$ or $W<W_c$, thus
\begin{equation}~\label{Eq:DoS_HOTDSM}
\rho(E)\sim \delta^{(d-z)\nu}(|E||\delta|^{-z\nu})^{d-1}=|E|^{d-1}|\delta|^{-(z-1)d\nu}.
\end{equation}
By contrast, inside the metallic phase the DOS at zero energy is finite, leading to
\begin{equation}~\label{Eq:DoS_metal}
\rho(E\approx 0)\sim \delta^{(d-z)\nu}(|E|\delta^{-z\nu})^0=\delta^{(d-z)\nu},
\end{equation}
for $\delta>0$ or $W>W_c$. Lastly, at the critical point ($\delta=0$) $\xi$ diverges, and therefore the $\xi$ independence of $G$ implies
\begin{equation}~\label{Eq:DoS_Critical}
\rho(E)\sim \delta^{(d-z)\nu}(|E|\delta^{-z\nu})^{(d-z)/z}=|E|^{(d-z)/z}.
\end{equation}
Since DOS at zero energy vanishes for $W \leq W_c$ and becomes finite in a metal, one can treat $\rho(0)$ as a bonafide order-parameter in dirty HOTDSM.

Numerically we reconstruct the average DOS by using the KPM in a cubic system of linear dimension $L=200$ for various choices of $t$ and $t_1$. First, we identify the critical disorder strength $W_c$, where $\rho(0)$ deviates from zero. Subsequently, we compute (a) the DSE $z$ from the scaling of $\rho(E)$ around $W=W_c$ [Eq.~(\ref{Eq:DoS_Critical})], and (b) the order-parameter exponent $\beta \equiv (d-z)\nu$ from the scaling of $\rho(0)$ inside the metallic phase [Eq.~(\ref{Eq:DoS_metal})]. Finally, from the known values of $z$ and $\beta$, we compute the CLE $\nu$. The results are summarized in Table~\ref{tab:exponents} and details of the numerical analysis are shown in the Appendix~\ref{Sec:DOS_Scaling}. Across a wide range of hopping parameters ($t$ and $t_1$) we find that $z\approx 1.5$ and $\nu\approx 1.0$ in a periodic system with linear dimension $L=200$ in each direction, which are fairly close (within the numerical accuracy) to the ones found for first-order Dirac and Weyl semimetals.

With the numerically extracted values of the critical exponents, we obtain convincing data collapses by comparing $\rho(E) |E|^{-\nu(d-z)}$ vs $E \delta^{-\nu z}$ in a periodic system with $L=200$, see Fig.~\ref{Fig:Collapse} (top row). All data points collapse onto three curves, corresponding to the DSM (lower ones), a metal (upper left ones) and the quantum critical regime (upper right ones). Finally, from the same set of exponents (computed in a $L=200$ periodic cubic system) we obtain excellent finite size data collapses by comparing $\rho(0) L^{d-z}$ with $L^{1/\nu} \delta$ inside the metallic phase for $100 \leq L \leq 200$ [Fig.~\ref{Fig:Collapse} (bottom row)]. Therefore, within a sufficiently wide range of system size $100 \leq L \leq 200$ the critical exponents ($\nu$ and $z$) do not change within our numerical accuracy. All together our extensive numerical analyses strongly suggest an emergent superuniversality across the DSM-metal QPT irrespective of their symmetries and the topological order (first or second), which now we substantiate from a leading-order momentum space RG analysis.

\section{Renormalization group analysis}~\label{sec:RG}

To perform the momentum space RG analysis in a dirty HOTDSM, we consider the following low-energy model
\begin{eqnarray}~\label{Eq:hDP}
\hat{h}_{\pm} (\vec{p})=v_\perp \sum_{j=1}^2 \Gamma_j p_j \pm v_3 p_3 \Gamma_3 + b \sum^5_{j=4} d_j(\vec{p})\Gamma_j,
\end{eqnarray}
obtained by expanding $\hat{h}^\vec{k}$ around one of the Dirac points at $\pm \vec{K}$ with $\vec{p}=-\vec{K}+\vec{k}$. Here $v_\perp=ta$, $v_3=t_z a$, $b=t_1 a^2/2$, and $(d_4,d_5)(\vec{p})=(p_2^2-p_1^2,2p_1 p_2)$. We retain the relevant higher-order momentum terms (proportional to $b$), such that $\hat{h}_\pm(\vec{p})$ is symmetry identical to the lattice models $\hat{h}^\vec{k}$, and represents the correct low-energy model for HOTDSM. The purpose of the following momentum space RG analysis is to provide a supportive argument in favor of the numerically observed superuniversality, resulting in (almost) identical values of the critical exponents $\nu$ and $z$ near the diffusive QCP in the entire family of DSMs that includes both first-order (for $b=0$) and higher-order (for $|b|>0$) DSMs.

Even though in the bare theory $v_3=v_\perp \equiv v$ (for $t=t_z$ in the lattice model), in general their RG flows are different when $|b|>0$, as it breaks the rotational symmetry between ${\bf p}_{_\perp}=(p_1,p_2)$ and $p_3$. To incorporate the effects of disorder we consider the following imaginary time ($\tau$) Eucledian action in $d$ dimensions~\cite{goswami-chakravarty-2, roy-slager-juricic-1}
\allowdisplaybreaks[4]
\begin{eqnarray}~\label{Eq:EuclS}
S&=&\int \D \tau \int \D^d x \; \Psi^\dag\Big[ \partial_\tau + \sum_{\alpha=\pm} \hat{h}_\alpha (\vec{p}\rightarrow -i {\boldsymbol \nabla}) - \Phi\Big]\Psi  \nonumber \\
&+&\frac{1}{2 \Delta}\int \D^d x \; \Phi \; |\boldsymbol{\nabla}|^{m} \; \Phi, 
\end{eqnarray}
where $\Phi$ is the disorder field that minimally couples to the four-component fermionic fields ($\Psi$) like a gauge field. The two-point correlator for the disorder fields in the real and momentum space are respectively 
\begin{eqnarray}~\label{Eq:Greenfn}
\langle \Phi(\vec{x})\Phi(\vec{y})\rangle=\frac{\Delta}{|\vec{x}-\vec{y}|^{d-m}}, \hspace{0.5cm}\langle \Phi(\vec{q})\Phi(\vec{0})\rangle=\frac{\Delta}{|\vec{q}|^{m}}.
\end{eqnarray}
As $m\rightarrow 0$, we recover Gaussian white noise distribution.

The scaling dimensions of momentum and frequency are $\left[q\right]=1$ and $\left[\omega\right]=z$, respectively. The scale invariance of $S$ implies $\left[\Psi\right]=d/2$, $\left[v\right]=z-1$, $\left[ b \right]=z-2$, and $\left[\Phi\right]=z+\eta_\Phi$, where $\eta_\Phi$ is the anomalous dimension of the disorder field, yielding $\left[\Delta\right]=2(z+\eta_\Phi)-(d-m)$. At the clean HOTDSM fixed point $z=1$ due to linearly dispersing excitations at sufficiently low energies, and $\eta_\Phi=0$ due to the gauge invariance of $S$. Therefore $\left[\Delta\right]=m-1$ in $d=3$, showing that (a) for Gaussian white noise distribution ($m=0$) disorder is an irrelevant perturbation at the HOTDSM fixed point (since $[\Delta]=-1$), and (b) the QPT to a metal at strong disorder can be addressed by performing a controlled RG analysis in terms of a \emph{small} parameter $\epsilon=1-m$, about $\epsilon=0$ for which disorder is \emph{marginal}, although ultimately we set $\epsilon=1$.

\begin{table}[t!]
\begin{tabular}{|c|c|c|c|c|c|c|c|c|c|}
\hline
$t_1$ & $t$ & $W_c$ & $\Delta W_c$& $\beta$ & $\Delta \beta$ & $z$ & $\Delta z$ & $\nu$ & $\Delta \nu$ \\
\hline
\multirow{5}{*}{0.0} & 0.50 & 2.8 & 0.1 & 1.47 & 0.03 & 1.50 & 0.059 & 0.98 & 0.05 \\
 & 0.75 & 3.4 & 0.1 & 1.48 & 0.02 & 1.52 & 0.053 & 1.00 & 0.05 \\
 & 1.00 & 3.9 & 0.1 & 1.47 & 0.02 & 1.49 & 0.05 & 0.98 & 0.04 \\
 & 1.25 & 4.4 & 0.1 & 1.43 & 0.02 & 1.50 & 0.044 & 0.95 & 0.04 \\
 & 1.50 & 4.8 & 0.1 & 1.41 & 0.03 & 1.50 & 0.027 & 0.94 & 0.03\\
\hline
\multirow{5}{*}{0.5} & 0.50 & 3.0 & 0.1 & 1.49 & 0.02 & 1.55 & 0.025 & 1.03 & 0.03 \\
 & 0.75 & 3.6 & 0.1 & 1.35 & 0.02 & 1.50 & 0.057 & 0.90 & 0.05 \\
 & 1.00 & 4.1 & 0.1 & 1.37 & 0.02 & 1.50 & 0.048 & 0.91 & 0.04 \\
 & 1.25 & 4.5 & 0.1 & 1.36 & 0.02 & 1.51 & 0.048 & 0.91 & 0.04 \\
 & 1.50 & 5.0 & 0.1 & 1.40 & 0.02 & 1.51 & 0.061 & 0.94 & 0.05\\
\hline
\multirow{5}{*}{1.0} & 0.50 & 3.3 & 0.1 & 1.63 & 0.03 & 1.50 & 0.008 & 1.08 & 0.02 \\
 & 0.75 & 3.9 & 0.1 & 1.48 & 0.02 & 1.51 & 0.046 & 0.99 & 0.04 \\
 & 1.00 & 4.4 & 0.1 & 1.58 & 0.02 & 1.45 & 0.026 & 1.02 & 0.03 \\
 & 1.25 & 4.7 & 0.1 & 1.53 & 0.03 & 1.48 & 0.022 & 1.00 & 0.03 \\
 & 1.50 & 5.2 & 0.1 & 1.55 & 0.04 & 1.48 & 0.008 & 1.02 & 0.03 \\
 \hline
 \multirow{5}{*}{1.5} & 0.50 & 3.6 & 0.1 & 1.63 & 0.03 & 1.51 & 0.004 & 1.09 & 0.02 \\
 & 0.75 & 4.1 & 0.1 & 1.62 & 0.03 & 1.50 & 0.002 & 1.08 & 0.02 \\
 & 1.00 & 4.6 & 0.1 & 1.51 & 0.02 & 1.50 & 0.025 & 1.01 & 0.03 \\
 & 1.25 & 5.1 & 0.1 & 1.46 & 0.02 & 1.51 & 0.021 & 0.98 & 0.03 \\
 & 1.50 & 5.5 & 0.1 & 1.55 & 0.04 & 1.50 & 0.017 & 1.03 & 0.04 \\
 \hline
\multirow{5}{*}{2.0} & 0.50 & 3.9 & 0.1 & 1.66 & 0.04 & 1.50 & 0.021 & 1.11 & 0.04 \\
 & 0.75 & 4.4 & 0.1 & 1.58 & 0.02 & 1.51 & 0.013 & 1.06 & 0.03 \\
 & 1.00 & 4.9 & 0.1 & 1.47 & 0.03 & 1.51 & 0.016 & 0.99 & 0.03 \\
 & 1.25 & 5.4 & 0.1 & 1.42 & 0.02 & 1.50 & 0.014 & 0.95 & 0.02 \\
 & 1.50 & 5.8 & 0.1 & 1.48 & 0.04 & 1.50 & 0.035 & 0.99 & 0.05 \\
 \hline
\multicolumn{4}{|c|}{Average} & 1.496 & 0.026 & 1.500 & 0.030 & 0.997 & 0.036\\
\hline
\end{tabular}
\caption{Summary of the scaling analysis for the DOS $\rho(E)$ in a cubic system of $L=200$. Here $t_1=0$ (finite) corresponds to first- (second-)order DSM. While the critical disorder ($W_c$) for metallicity (a nonuniversal quantity) depends on the hopping parameters, $t$ and $t_1$ [Eq.~(\ref{Eq:tb})], the critical exponents $z$ and $\nu$ are centered around $1.5$ and $1.0$, respectively. This observation suggests that the universality class of the DSM-metal QPT does not depend on the symmetry or topological order (first or second), thereby yielding a superuniversality in the entire family of disordered DSM, in qualitative agreement with the findings from the momentum space RG analysis. Here $\Delta X$ is the fitting error of $X$, for $X=z,\beta,\nu$. For details and further data see Appendix~\ref{Sec:DOS_Scaling} and the supplementary material~\cite{SM}.
}~\label{tab:exponents}
\end{table}

To derive the RG flow equations, we integrate out a thin momentum shell $\left[ \Lambda e^{-\ell}, \Lambda \right]$, where $\ell(>0)$ is the logarithm of the RG scale and $\Lambda$ is the ultraviolet cutoff. The relevant Feynman diagrams are shown in Fig.~\ref{Fig:fig4}(c) and (d). After accounting for the quantum corrections up to the leading-order, the RG flow equations read (see Appendix~\ref{Sec:RG_details} for details)
\allowdisplaybreaks[4]
\begin{eqnarray}~\label{Eq:RGflow}
\frac{\D v_3}{\D \ell} &=& g v_3 \left[ 2 f_1(x)-f_3(x) \right] \equiv (z-1)v_3, \nonumber  \\ 
\frac{\D g}{\D \ell} &=& g \left[ -\epsilon + 2(z-1)\right], \:\:
\; \frac{\D x}{\D \ell}=-x + {\mathcal O} (x g), \nonumber \\ 
\frac{\D}{\D \ell}\left(\frac{v_\perp}{v_3}\right)&=&g\left(\frac{v_\perp}{v_3}\right) \left[f_3(x)-f_\perp(x)\right],~
\end{eqnarray}
where $g=\Delta \Lambda^{\epsilon}/(2 \pi^2 v^2)$ is the dimensionless disorder coupling, $x=b \Lambda/v$ is also dimensionless, and 
\allowdisplaybreaks[4]
\begin{eqnarray}~\label{Eq:RGfuncs}
f_1(x)&=&\frac{1}{2}\int_0^\pi \D \theta \frac{S_\theta}{1+x^2 S^4_\theta},
f_3(x)=\int_0^\pi \D \theta \frac{S_\theta C^2_\theta}{(1+x^2 S^4_\theta)^2}, \nonumber \\
f_{\perp}(x)&=&\frac{1}{2}\int_0^\pi \D \theta \frac{ S^3_\theta \; (1+2x^2 S^2_\theta)}{(1+x^2 S^4_\theta)^2},
\end{eqnarray}
with $S_\theta \equiv \sin \theta$, $C_\theta \equiv \cos \theta$. Since we are interested in the leading-order RG analysis, (1) quantum corrections are computed by setting $v_\perp=v_3=v$, and (2) only the engineering dimension of $x$ has been taken into account. The flow equation of $v_3$ is fixed by its scaling dimension, yielding dynamic scaling exponent $z=1+g\left[2 f_1(x)-f_3(x)\right]$. The flow equation for $g$ then supports two fixed points: (1) an infrared stable one at $g=0$, describing a clean HOTDSM, and (2) an infrared unstable QCP at $g=g_\ast=\epsilon/[2(2f_1(x)-f_3(x))]$. The latter one controls the QPT into a metal at finite disorder, where $z=1+\epsilon/2=3/2$ for Gaussian white noise distribution ($\epsilon=1$) for any $x$. The CLE, defined as $\nu^{-1}=\D(\D g/\D \ell)/\D g|_{g=g_\ast}=\epsilon$ at the dirty QCP. Note that $(g_\ast, x)$ determines the phase boundary between the DSMs and a metal [solid blue line in Fig.~\ref{Fig:fig1}(b)], which is symmetric about $x=0$, as all the functions in Eq.~(\ref{Eq:RGfuncs}) are symmetric under $x \to -x$. One also obtains \emph{identical} DSE $z$ from the flow equation of $v_\perp$ (see Appendix~\ref{Sec:RG_details}).

Finally we note that the four-fold symmetry breaking Wilson mass ($x$) is always an \emph{irrelevant} parameter. Thus in the deep infrared regime ($\ell \to \infty$) $x \to 0$, where $f_\perp(0)=f_3(0)=2/3$, and the velocity anisotropy becomes \emph{marginal}. Consequently, the ratio $v_\perp/v_3$ ultimately flows to its bare value, set by the hopping parameters $t$ and $t_z$. Therefore, the Wilson mass ($x$) despite breaking the time-reversal and $C_4$ symmetries only changes the location of the dirty QCP ($g_\ast$) without altering the universality class of the semimetal-metal QPT (determined by $\nu$ and $z$, see Table~\ref{tab:exponents}), giving rise to a \emph{superuniversality} in the entire family of dirty DSMs, that includes its first- and second-order cousins. Although not guaranteed a priori, the irrelevance of $x$ does not change even after we account for its entire one-loop quantum correction (see Appendix~\ref{Sec:RG_details}), which is in agreement with the exact numerical findings that the universality class the HOTDSM-metal QPT remains unchanged even when $t_1 \gg t$ [see Table~\ref{tab:exponents}].

\section{Summary and Discussion}~\label{sec:summary}

Here we investigate the stability of a HOTDSM in the presence of quenched charge impurities, and identify a semimetal-metal QPT at finite disorder. While the topological hinge modes gradually melt across this transition in open boundary systems [Fig.~\ref{Fig:FadingHinge}], similar to the Fermi arcs in dirty Weyl semimetals~\cite{roy-slager-juricic-2}, we come to the conclusion that the symmetry and topological order (first or second) does not affect its universality class [Table~\ref{tab:exponents}], thus yielding a \emph{superuniversality} in the entire family of dirty DSMs. Since real materials are inherently dirty, the stability of HOTDSM is critical for its experimental realization and observation of the hinge modes (via STM and ARPES, for example) in sufficiently clean systems. We also note that HOTDSM in general is more stable than its first-order counterpart due to the suppression of DOS by the Wilson mass [Fig.~\ref{Fig:fig1}]. In the future, it will be worthwhile to investigate the nature of the critical wavefunctions in dirty HOTDSM, and search for measurable signatures of the discrete $C_4$ symmetry breaking.

\acknowledgments

B.R. was supported by the Startup grant from Lehigh University. 

\appendix
\widetext

\section{Renormalization group analysis}~\label{Sec:RG_details}

In this appendix we present the RG analysis of disordered HOTDSM in details. 
In what follows we focus on one of the Dirac points located at $-\vec{K}$ and take $\hat{h}_+({\bf p}) \to \hat{h}({\bf p})$ in Eq.~(\ref{Eq:hDP}) for notational compactness. Such a simplification does not alter any outcome as disorder does not couple two valleys and therefore they can be treated as independent flavors. Moreover, the flavor number does not affect the RG flow equations as the perturbative RG analysis does not involve any Feynman diagram that contains fermion bubble or the flavor number~\cite{roy-dassarma-1}. 
The low energy Hamiltonian around the Dirac point at $-\vec{K}$ reads [see the lattice model in Eq.~(\ref{Eq:tb})]
\begin{equation}~\label{Eq:hDP_2}
\hat{h}({\vec{p}})=v_\perp \sum_{j=1}^2 \Gamma_j p_j + v_3 p_3 \Gamma_3 + b(p_2^2-p_1^2)\Gamma_4 +b (2p_1 p_2) \Gamma_5,
\end{equation}
where ${\bf K}=\left[0,0,\pi/2\right]$ and ${\bf p}=-{\bf K}+{\bf k}$. Even though in the lattice model we set the bare Fermi velocities to be equal, i.e. $v_\perp^\mathrm{bare}=v_3^\mathrm{bare}$ (by setting $t=t_z$), due to the $C_4$ symmetry breaking terms (proportional to $b$) they receive different quantum corrections, as the rotational symmetry between ${\bf p}_\perp=(p_1,p_2)$ and $p_3$ is broken when $b \neq 0$. The Euclidean action in the presence of disorder reads as
\begin{equation}
S = \int \D \tau \D^3 x\Psi^\dag\Big[ \partial_\tau + \hat{h}({\vec{p}\rightarrow -i \boldsymbol{\nabla}}) - \Phi\Big]\Psi + \frac{1}{2\Delta} \; \int \D^3 x \Phi |\boldsymbol{\nabla}|^m \Phi, \label{Eq:EuclS}
\end{equation}
where $\Psi$ is the fermion field and $\Phi$ represents the (bosonic) disorder field. The fermionic and disorder Green's functions are respectively given by
\begin{eqnarray}
G(i\omega,\vec{p})&=&-\frac{ i\omega+ \hat{h}({\bf p})}{\omega^2+v^2 p^2+b^2 p_\perp^4} 
\hspace{0.7cm}\mathrm{and} \hspace{0.7cm}
\langle \Phi(\vec{p})\Phi(\vec{0})\rangle=\frac{\Delta}{|\vec{p}|^{m}}.
\end{eqnarray}
While computing the leading order quantum corrections due to disorder, see Fig.~\ref{Fig:fig4} bottom row, we set $v_\perp=v_3=v$ in the fermionic Green's function.

In the Wilsonian RG procedure, we integrate out a thin momentum shell $\left[ \Lambda e^{-\ell}, \Lambda \right]$, with $0< \ell \ll 1$, where $\Lambda$ is the ultraviolet cutoff. Subsequently, we recast the action ($S$) in its original form, but in terms of the renormalized or scale ($\ell$) dependent quantities, leading to their RG flow equations. We begin with determining the engineering dimensions of various quantities in Eq.~(\ref{Eq:EuclS}). Since momentum and frequency respectively scale as $\left[q\right]=1$ and $\left[\omega \right]=z$, where $z$ is the dynamic scaling exponent, the scaling dimensions for their conjugate variables are $\left[x\right]=-1$ and $\left[\tau \right]=-z$. The scale invariance of the Euclidean action then implies $\left[ \Psi \right]=d/2$, $\left[v\right]=z-1$ and $\left[ b \right]=z-2$.

Note that the $\Phi$ field appears in Eq.~(\ref{Eq:EuclS}) exactly as $i\omega$, consequently at the clean fixed point it bears the same scaling dimension as $1/\tau$, so $\left[\Phi \right]=z+\eta_\Phi$, where we introduced an anomalous dimension of the disorder field $\eta_\Phi$, to allow for possible quantum corrections arising at the disorder controlled fixed point. Imposing the above relations on the second term of Eq.~(\ref{Eq:EuclS}) we obtain
\begin{equation}
\left[ \Delta \right] = 2(z+\eta_\Phi)-(d-m).
\end{equation}
At the clean fixed point (describing a HOTDSM) $z=1$ for linearly dispersing fermions at sufficiently low energies, yielding $\eta_\Phi=0$. Notice that the disorder field couples with the fermionic field like a gauge field. Consequently, $\eta_\Phi$=0 always, due to the Ward identity, as we demonstrate below for the leading-order RG analysis. In three dimensions $\left[ \Delta \right] = m-1$. As seen in Eq.~(\ref{Eq:Greenfn}), $m=0$ corresponds to Gaussian white noise distribution, for which disorder is irrelevant in the language of RG. On the other hand, for $m=1$ disorder becomes marginal, which motivates our $\epsilon$ expansion in terms of a small parameter $\epsilon=1-m$. Accordingly we define a dimensionless disorder coupling
\begin{equation}
g=\frac{\Delta\Lambda^{\epsilon}}{2 \pi^2 v^2},
\end{equation}
which to leading order scales identically as $\Delta$. Having established the scaling of various quantities appearing in the action $S$, we proceed with evaluating the two relevant one-loop Feynman diagrams in Fig.~\ref{Fig:fig4}(c) and (d), namely the fermionic self energy (c) and the correction to the fermion-disorder vertex (d).

\subsection{Fermionic self energy}

First we compute the contribution from the self energy diagram, given by 
\begin{equation}
\Sigma (i\omega, \vec{k})= \Delta \int \frac{\D^3 \vec{q}}{(2\pi)^3} G(i\omega, \vec{k}-\vec{q}) \frac{1}{|q|^m}.~\label{Eq:Sigma(w,k)}
\end{equation}
Since to the leading-order the divergences in the spatial and temporal part of the self energy are separated (no overlapping divergences), we evaluate $\Sigma (i\omega, 0)$ and $\Sigma (0, \vec{k})$ separately, and $\Sigma (i\omega, \vec{k})=\Sigma (i\omega, 0)+\Sigma (0, \vec{k})$. Let us write down the time-like component
\begin{align}
\Sigma(i\omega,0)=-\Delta (i\omega)\int\frac{\D^3\vec{q}}{(2\pi)^3}\frac{|q|^{-m}}{\omega^2+v^2 q^2+b^2 q^4 \sin^4\theta}.
\end{align}
Since we compute this contribution within the Wilsonian momentum shell $[\Lambda e^{-\ell}, \Lambda]$, we set $\omega=0$ in the integrand, leading to 
\begin{align}
\Sigma(i\omega,0)=-\frac{\Delta(i\omega)}{4 \pi^2} \int^{\Lambda}_{\Lambda e^{-\ell}} \D q \int^\pi_0 \D \theta \frac{ q^{2-m} \sin \theta}{v^2 q^2+b^2 q^4 \sin^4 \theta}
=-(i\omega) g l f_1(x),
\end{align}
where $x=\Lambda b/v$ and $f_1(x)$ has been defined in Eq.~(\ref{Eq:RGfuncs}).

Next we set $\omega=0$ and evaluate $\Sigma (0, \vec{k})$, that to the $k$-linear order renormalizes the Fermi velocities $v_\perp$ and $v_3$, and to the quadratic order in momentum gives corrections to $b$. By setting ${\bf k}=\left[k_x,0,0\right]$, we collect the renormalization for both $v_\perp$ and $b$, while for ${\bf k}=\left[0,0,k_z\right]$ only $v_3$ gets renormalized. Hence, we evaluate $\Sigma (0, \vec{k})$ for these two choices of ${\bf k}$ separately. Let us begin by taking $\vec{k}= \left[k_x,0,0\right]$ for which 
\begin{equation}
\Sigma(0,\vec{k})=-\Delta \int \frac{\mathrm{d}^3 \vec{q}}{(2 \pi)^3}
\frac{v \left[(k_x-q_x) \Gamma_1-q_y \Gamma_2-q_z \Gamma_3 \right]+b \left[ (k_x-q_x)^2-q_y^2\right] \Gamma_4-2b (k_x-q_x)q_y \Gamma_5}
{v^2(k_x^2+q^2-2 k_x q_x)+b^2\left[ (k_x-q_x)^2-q_y^2 \right]^2+4 b^2(k_x-q_x)^2 q_y^2} \;
\frac{1}{|q|^{m}},\label{Eq:Sigma(kx)}
\end{equation}
and expand the denominator up to (for now) the linear order in $k_x$ as
\begin{equation}
\mathrm{denom}(k_x)=
\frac{1}{v^2 q^2 + b^2 q_\perp^4}+
\frac{ 2 q_x \left[2 b^2 q_\perp^2 + v^2\right] k_x}{\left[v^2 q^2 + b^2 q_\perp^4 \right]^2}+ \mathcal{O}(k_x^2).
\end{equation}
Substituting the above expansion into Eq.~(\ref{Eq:Sigma(kx)}) and performing the $\phi$ integral yields
\begin{equation}
\Sigma(0,\vec{k})= -v_\perp k_x \Gamma_1 \; \frac{\Delta}{4 \pi^2} \int^\Lambda_{\Lambda e^{-\ell}} \D q   q^2 \int^\pi_0 \D \theta \sin \theta \Bigg( \frac{1}{v^2 q^2 + b^2 q_\perp^4} - 
\frac{v q_\perp^2(v^2+2 b^2 q_\perp^2)}{(v^2 q^2 + b^2 q_\perp^4)^2} \Bigg)
= -v_\perp gl\left[f_1(x)-f_\perp(x) \right] \Gamma_1 k_x,
\end{equation}
where $f_\perp (x)$ is introduced in Eq.~(\ref{Eq:RGfuncs}).

Because of the $b(k_x^2-k_y^2)$ term in Eq.~(\ref{Eq:hDP_2}), we continue the above expansion to $k_x^2$ order, which will yield quantum corrections to $b$. For brevity we do not write out the $\mathcal{O}(k_x^2)$ term in the Taylor expansion, rather quote the (significantly simpler) result after the $\phi$ integration
\begin{eqnarray}
-bk_x^2 &\Gamma_4 & \frac{\Delta}{4 \pi^2} \int^\Lambda_{\Lambda e^{-\ell}} \D q q^2 \int^\pi_0 \D \theta  \sin \theta \Bigg( \frac{1}{v^2 q^2+b^2 q_\perp^4}-2 \frac{q_\perp^2(v^2+ 2 b^2 q_\perp^2)}{(v^2 q^2+b^2 q_\perp^4)^2} + \frac{q_\perp^4 \left[ v^4+3b^4q_\perp^4-v^2b^2(q^2-4q_\perp^2)\right]}{(v^2 q^2+b^2 q_\perp^4)^3} \Bigg) \frac{1}{|q|^{m}}\nonumber \\
& =&-bgl \left[ f_1(x)-2 f_\perp(x)+f_b(x)\right] \Gamma_4 k_x^2,
\end{eqnarray}
where we introduce yet another function
\begin{equation}
f_b(x)=\frac{1}{2}\int_0^\pi \D \theta \; \frac{1+3x^4 \sin^4\theta-x^2(1-4 \sin^2 \theta)}{(1+x^2 \sin^4\theta)^3} \; \sin^5\theta.
\end{equation} If we now take $\vec{k}=[0,k_y,0]$, we obtain the above correction with an overall negative sign, corresponding to $b(k_x^2-k_y^2)$.

As the next step let us take $\vec{k}= \left[0,0,k_z\right]$, after which Eq.~(\ref{Eq:Sigma(w,k)}) becomes
\begin{equation}
\Sigma(0,\vec{k})= \int \frac{\D^3 \vec{q}}{(2\pi)^3}
\frac{v \left[-q_x \Gamma_1-q_y \Gamma_2+(k_z-q_z) \Gamma_3 \right]+b (q_x^2-q_y^2) \Gamma_4+2 b q_x q_y \Gamma_5}
{v^2(k_z^2+q^2-2 k_z q_z)+b^2 (q_x^2-q_y^2)^2+4 b^2q_x^2 q_y^2} \: \frac{\Delta}{|q|^m},
\end{equation}
and expand the denominator in $k_z$ to linear order as
\begin{equation}
\mathrm{denom}(k_z)=\frac{1}{ v^2 q^2 + b^2 q_\perp^4}+\frac{2 v^2 q_z k_z}{\left[v^2q^2 + b^2 q_\perp^4 \right]^2}+ \mathcal{O}(k_z^2),
\end{equation}
and like before, substitute it into $\Sigma(0,\vec{k})$, which leads to 
\begin{equation}
\Sigma(0,\vec{k})=-v_3 k_z \Gamma_3 \; \frac{\Delta}{4 \pi^2} \int^\Lambda_{\Lambda e^{-\ell}} \D q q^2 \int^\pi_0 \D \theta \sin \theta \Bigg( \frac{1}{v^2 q^2 + b^2 q_\perp^4} - 
\frac{2 v^3 q_z^2}{(v^2 q^2 + b^2 q_\perp^4)^2} \Bigg) \frac{1}{|q|^m}
= -v_3gl\left[f_1(x)-f_3(x) \right] \Gamma_3 k_z.
\end{equation}
Since $k_z^2$ does not appear in the action we have no reason to proceed to this order in the Taylor expansion.

\subsection{Vertex correction}

Next we evaluate the vertex diagram, see Fig.~\ref{Fig:fig4}(d). The leading divergence of this diagram can be extracted by setting the external frequency to zero. Then the contribution reads
\begin{equation}
V(\vec{k})=\int \frac{\D^3 {\bf q}}{(2\pi)^3} \frac{\Gamma_0 \hat{h} (\vec{k}-\vec{q}) \Gamma_0 \hat{h} (\vec{k}-\vec{q})\Gamma_0}{ \left[ v^2 (\vec{k}-\vec{q})^2 + b^2 (\vec{k}_\perp -\vec{q}_\perp)^2 \right]^2} \:\: \frac{\Delta}{|q|^{m}}.
\end{equation}
We now evaluate the integral for $\vec{k}=0$, and obtain 
\begin{align}
V(0)=\frac{\Delta}{4 \pi^2} \int \frac{\D q \D \theta q^2 \sin\theta}{v^2 q^2+b^2 q_\perp^4} |q|^{-m}=g f_1(x)l \Gamma_0 
\equiv -\Sigma(i \omega, 0).
\end{align}
Therefore, the disorder field $(\Phi)$ does not receive any anomalous dimension ($\eta_\Phi$) from quantum corrections, according to the Ward identity.

\subsection{Flow equations}

Let us write the dressed fermion propagator as 
\begin{eqnarray}
G^{-1}(i\omega,\vec{k}) = G_0^{-1}(i\omega,\vec{k})+\Sigma(i\omega,\vec{k}) 
=Z_\Psi\Big[i\omega -Z_{v_\perp} v_\perp \sum_{j=1}^2 \Gamma_j k_j - Z_{v_3} v_3 \Gamma_3 k_3 - Z_b b \Big((k_x^2-k_y^2)\Gamma_4+2k_xk_y\Gamma_5\Big)\Big],
\end{eqnarray}
from where we read off the following renormalization conditions
\begin{eqnarray}
Z_\Psi &=& 1-g f_1(x) \ell + {\mathcal O}(\ell^2), \\
Z_{v_3} &=& Z_\Psi^{-1} \Big[ 1+g\Big( f_1(x)-f_3(x) \Big) \ell \Big]\approx 1+g\Big( 2f_1(x)-f_3(x) \Big) \ell + {\mathcal O}(\ell^2), \\
Z_{v_\perp} &=& Z_\Psi^{-1} \Big[ 1+g\Big( f_1(x)-f_\perp(x) \Big) \ell \Big]\approx 1+g\Big(2 f_1(x)-f_\perp (x) \Big)\ell + {\mathcal O}(\ell^2), \\
Z_b &=& Z_\Psi^{-1} \Big[ 1+g\Big( f_1(x)-2 f_\perp(x)+f_b(x) \Big)\ell \Big] \approx 1+g\Big( 2 f_1(x)-2 f_\perp(x)+f_b(x) \Big)\ell + {\mathcal O}(\ell^2).
\end{eqnarray}
Since $\Phi$ appears in Eq.~(\ref{Eq:hDP_2}) exactly the same way as $i\omega$, we can write analogously
\begin{equation}
Z_\Phi=Z_\Psi^{-1} \Big[1-gf_1(x) \ell \Big] = 1.
\end{equation}
In other words, to the leading order $g$ cancels in the renormalization of the disorder field, and $\eta_\Phi=0$. From $Z_b$ we can write down the renormalization condition for the corresponding dimensionless quantity $x=b \Lambda/v$ as 
\begin{equation}
Z_x= Z_b \; Z_\Lambda= 1 - \ell +g\Big( 2 f_1(x)-2 f_\perp(x)+f_b(x) \Big)\ell + {\mathcal O}(\ell^2),  
\end{equation}
since $Z_\Lambda =(1-\ell)$, as $\Lambda \to \Lambda e^{-\ell}$ under rescaling.

The RG flow equations are then given by 
\begin{eqnarray}
\frac{\D v_\perp}{\D \ell}&=&gv_\perp \Big( 2f_1(x)-f_\perp(x) \Big), \\
\frac{\D v_3}{\D \ell}&=&gv_3\Big( 2f_1(x)-f_3(x) \Big), \\
\frac{\D}{\D \ell}\frac{v_\perp}{v_3}&=&\frac{1}{v_3}\frac{\D v_\perp}{\D \ell}-\frac{v_\perp}{v_3^2}\frac{\D v_3}{\D \ell}=g\frac{v_\perp}{v_3} \Big(f_3(x)-f_\perp (x) \Big), \\
\frac{\D g}{\D \ell} &=& g \left[ -\epsilon + 2(z-1)\right] \\
\frac{\D x}{\D \ell}&=&-x\left[1 + g\Big( 2 f_1(x)-2f_\perp (x) +f_b (x) \Big) \right].
\end{eqnarray}
Since we are interested in the leading-order corrections to all variables, we only consider the contribution of the first term in the flow equations of $x$, and neglect the quantum corrections arising from disorder in Sec.~\ref{sec:RG}. 

In Section~\ref{sec:RG} we anayzed the above flow equations by setting $dv_3/d\ell=(z-1) v_3$. Alternatively, we can choose $dv_\perp/d\ell=(z-1) v_\perp$. Then $z=1+ g [2f_1(x)-f_\perp(x)]$ and the dirty QCP is located at $g=g_\ast=\epsilon/[2(2f_1(x)-f_\perp(x))]$, where $z=1+\epsilon/2$ and $\nu^{-1}=\epsilon$. Therefore, only the location of the dirty QCP $g_\ast$ changes (however, without altering its variation with $x$, qualitatively), which is a nonuniversal quantity. But the universal critical exponents, $z$ and $\nu$, are insensitive to these details.

\begin{figure}
\includegraphics[width=0.24\linewidth]{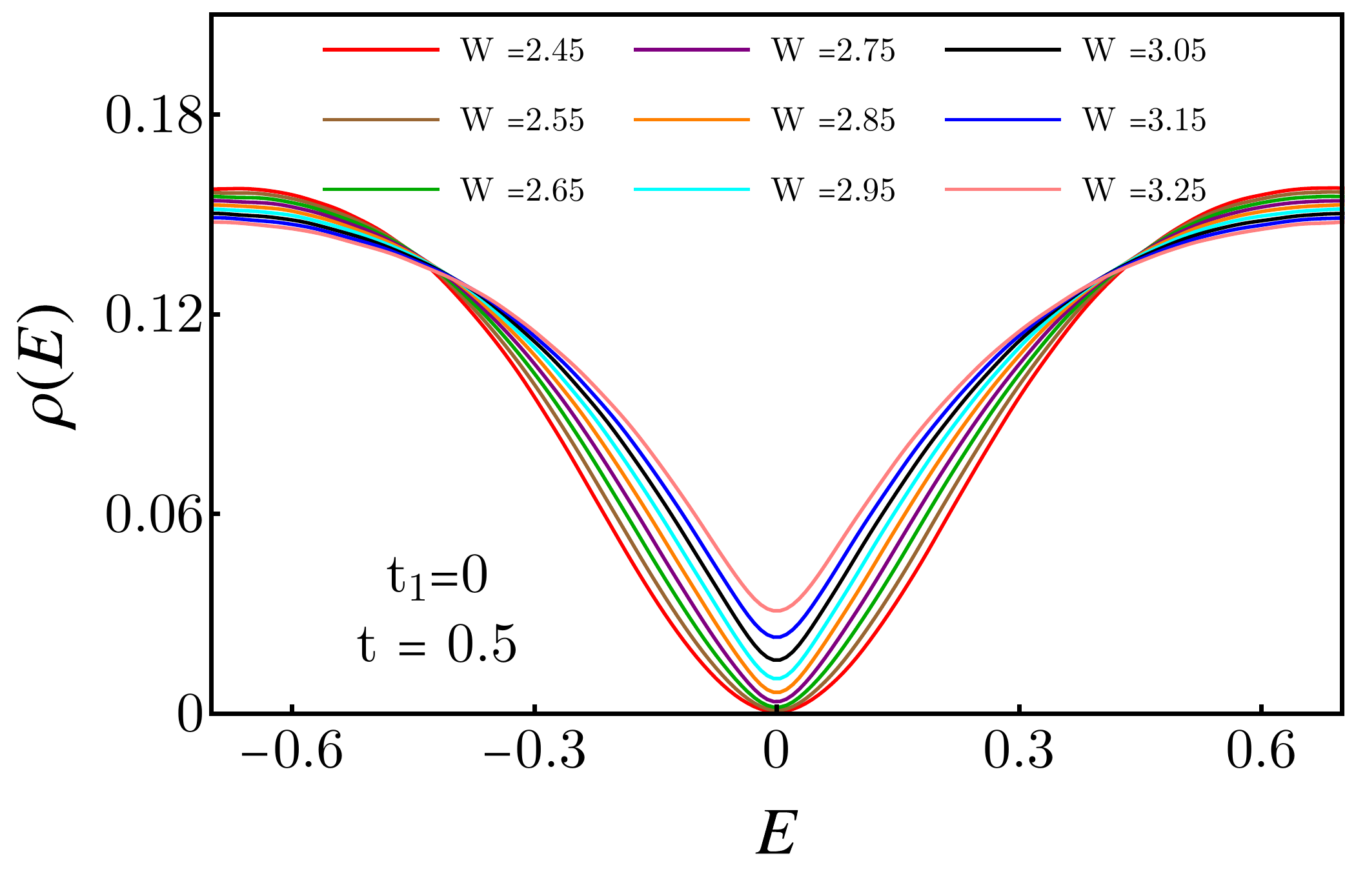}
\includegraphics[width=0.24\linewidth]{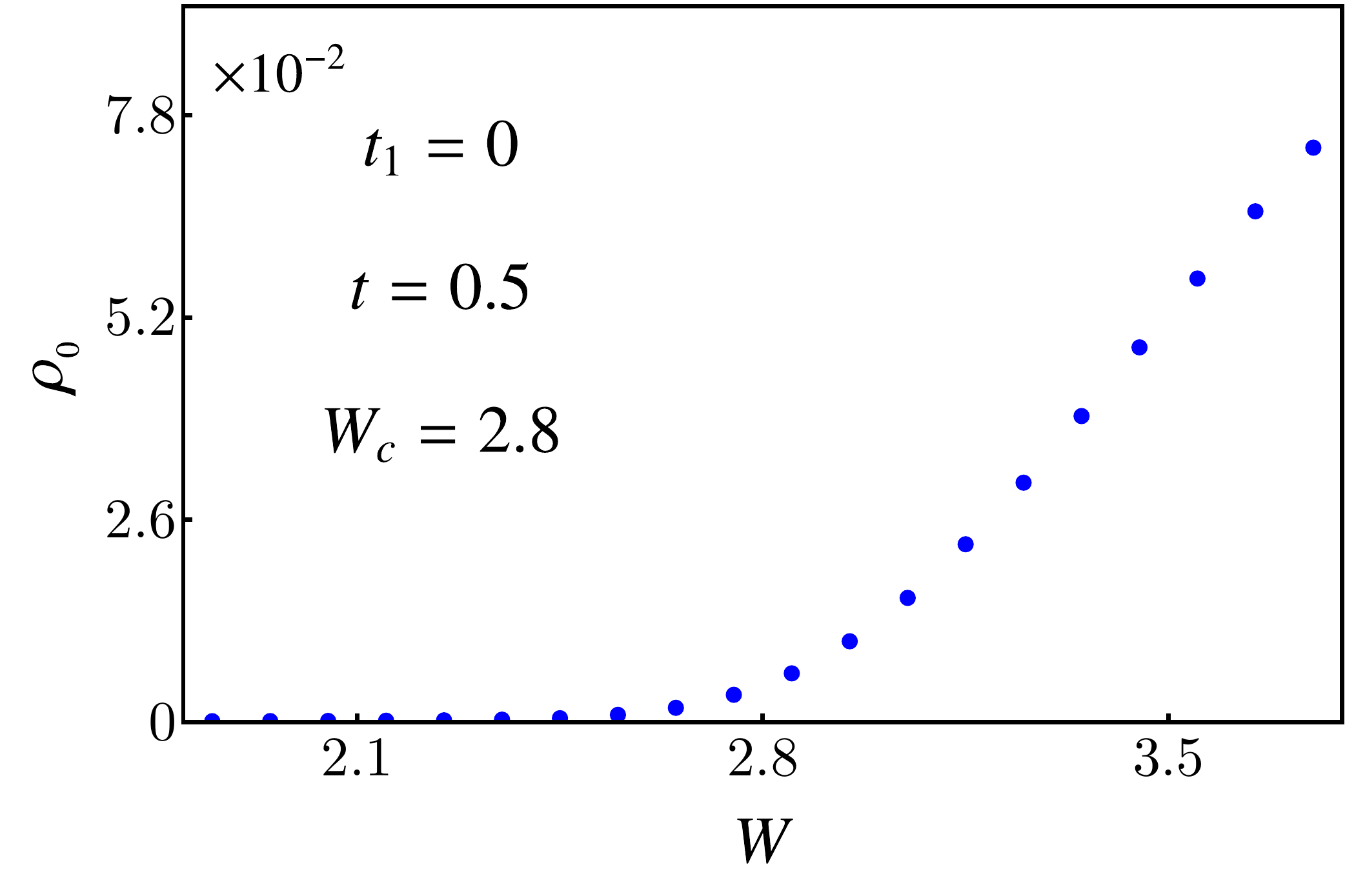}
\includegraphics[width=0.24\linewidth]{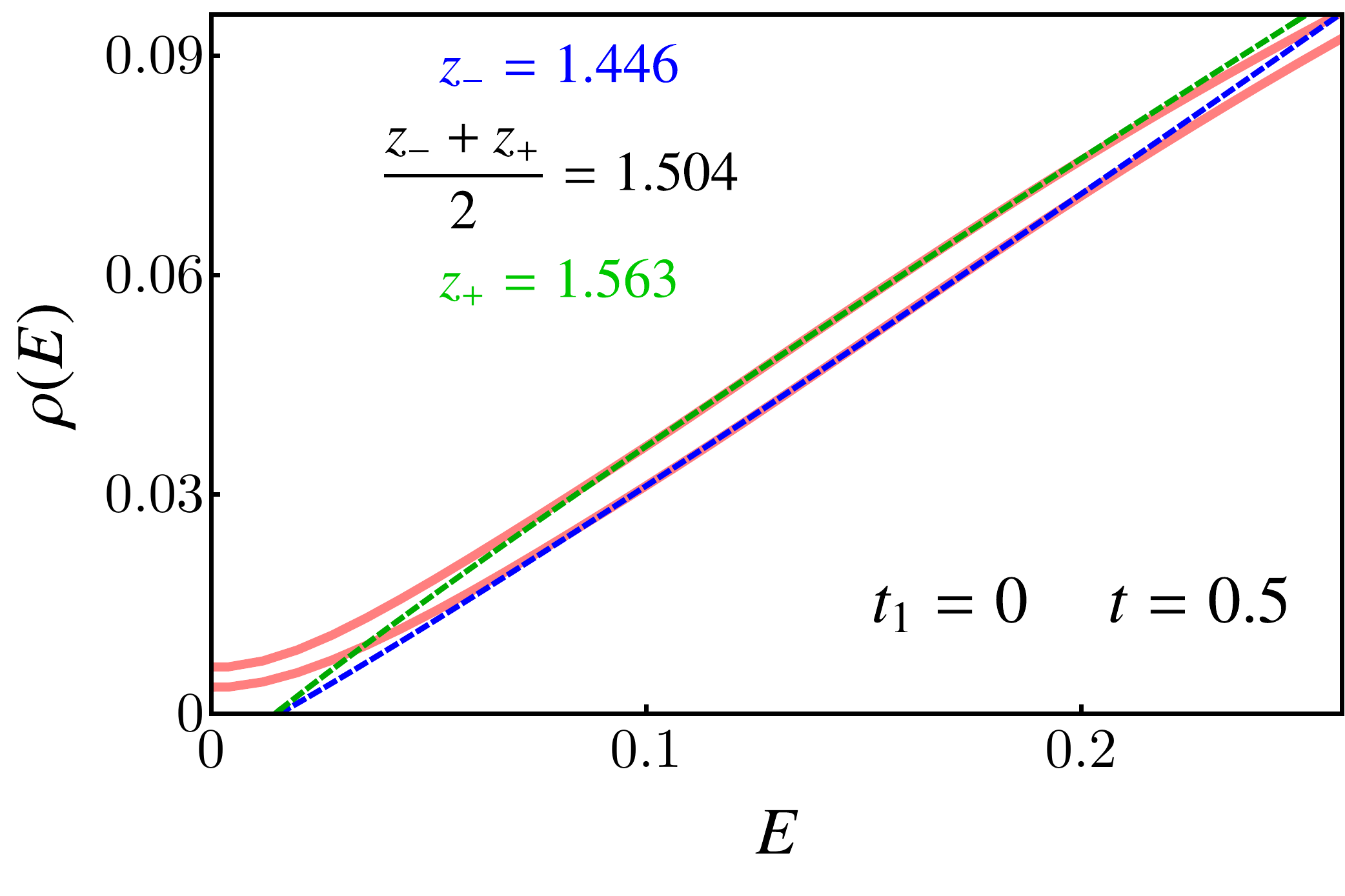}
\includegraphics[width=0.24\linewidth]{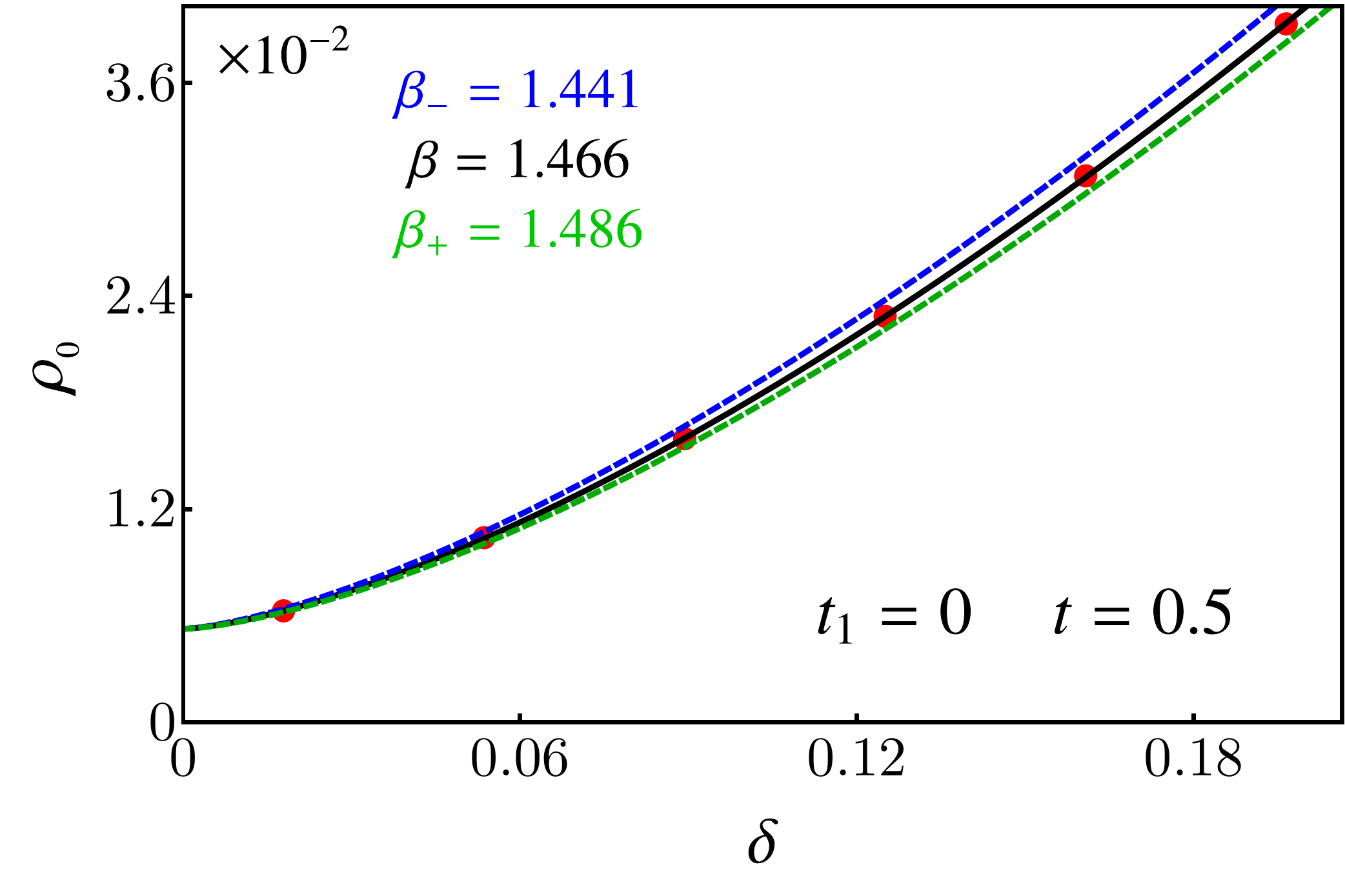}\\
\includegraphics[width=0.24\linewidth]{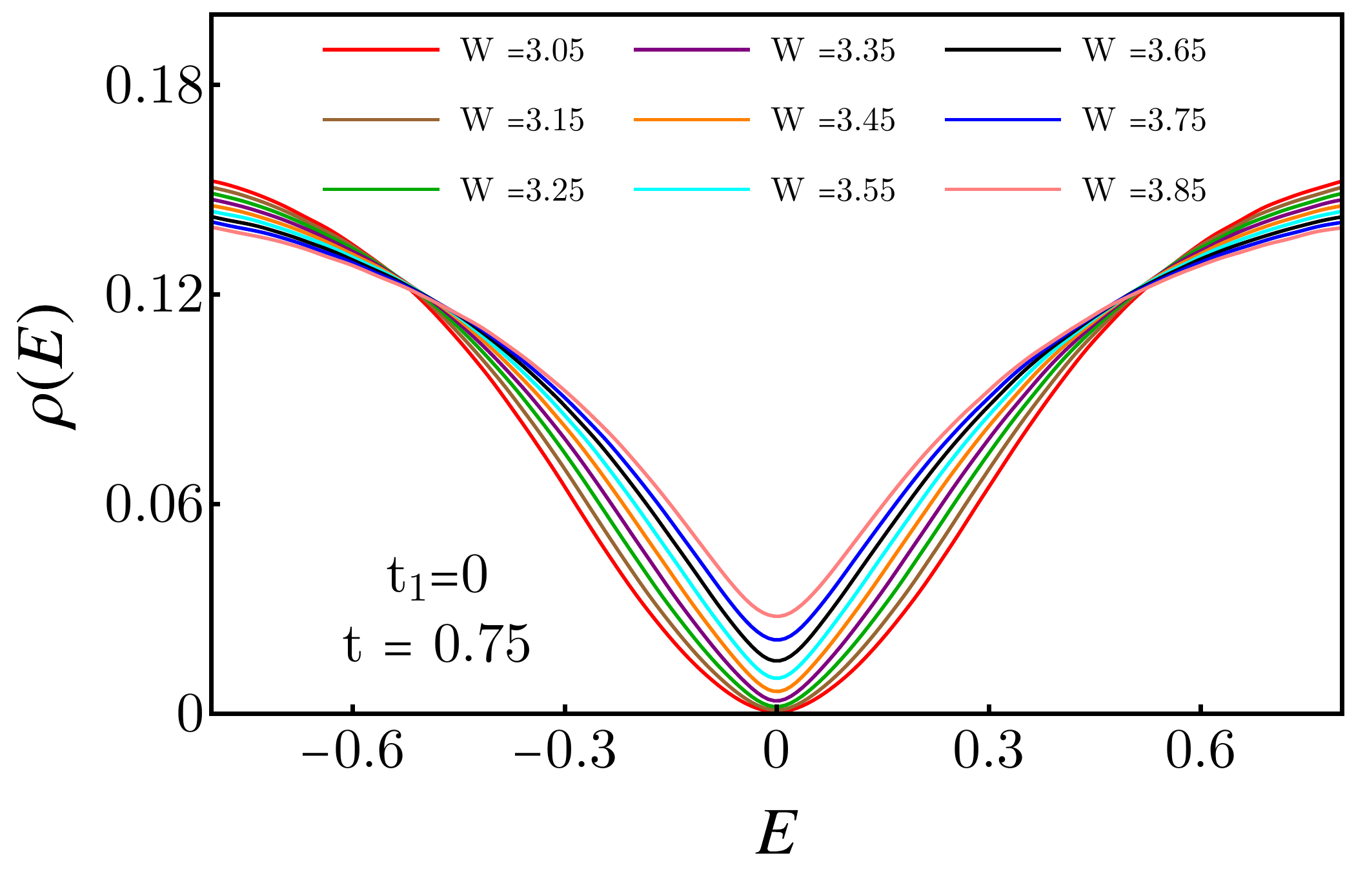}
\includegraphics[width=0.24\linewidth]{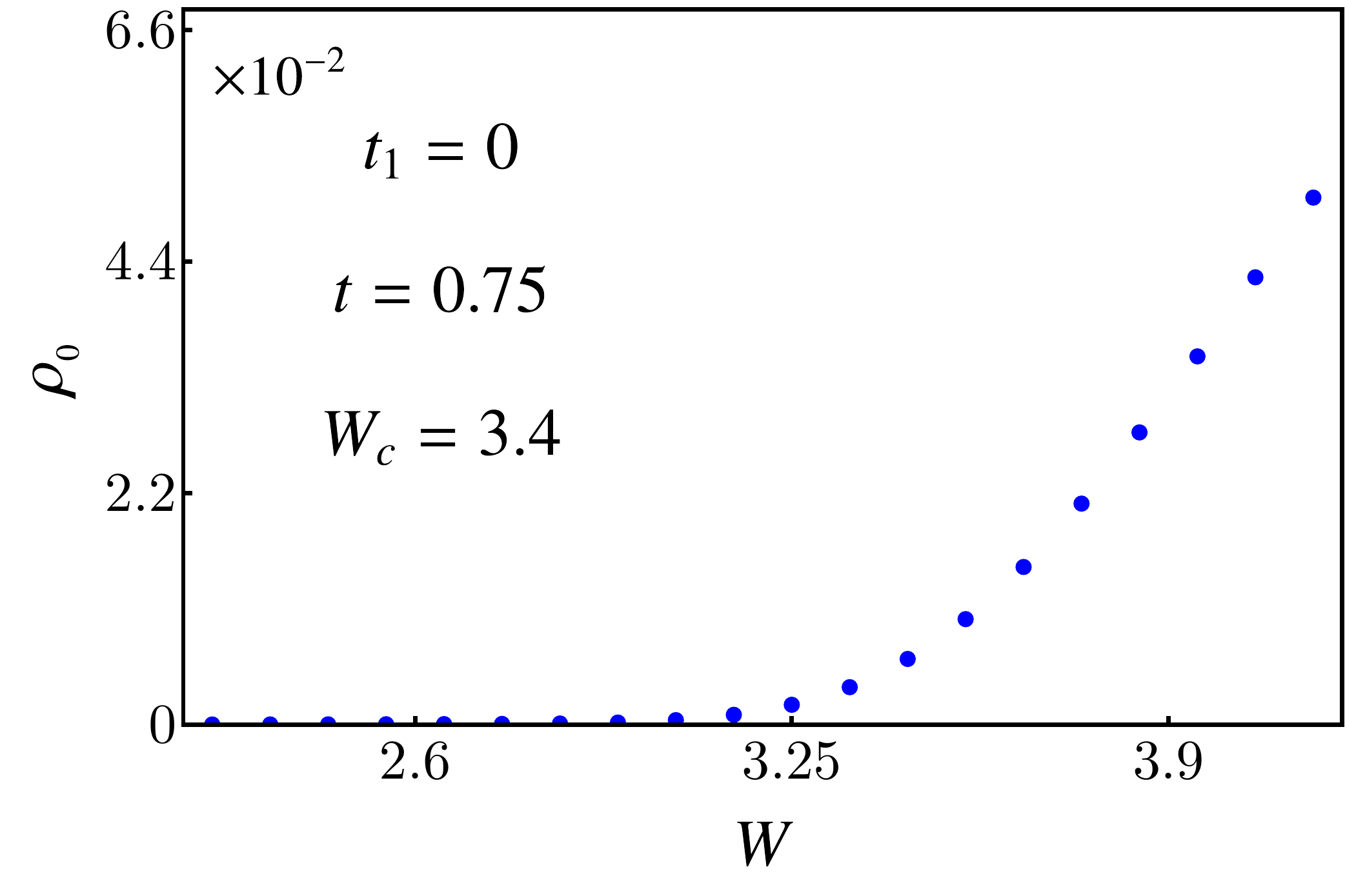}
\includegraphics[width=0.24\linewidth]{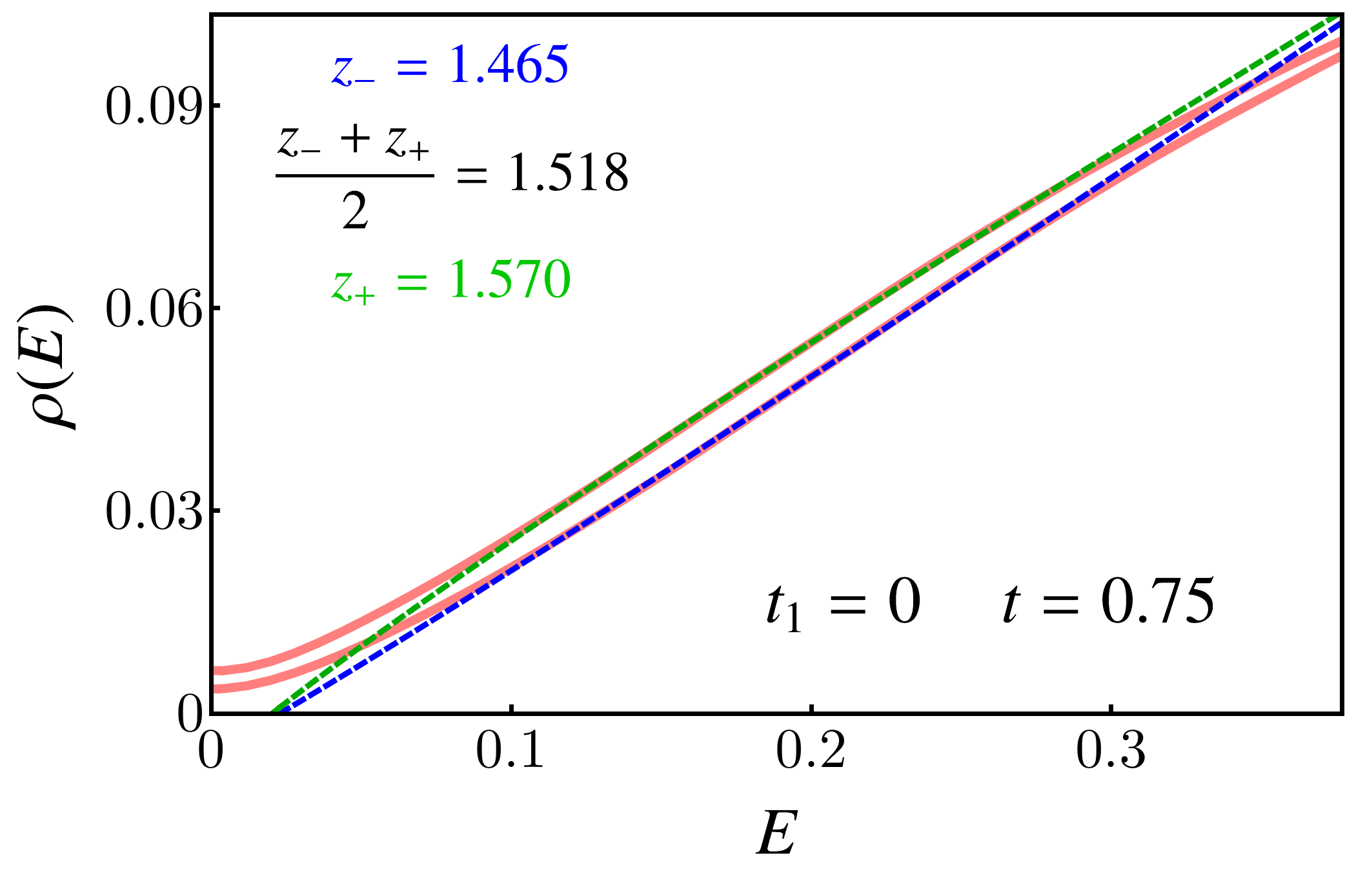}
\includegraphics[width=0.24\linewidth]{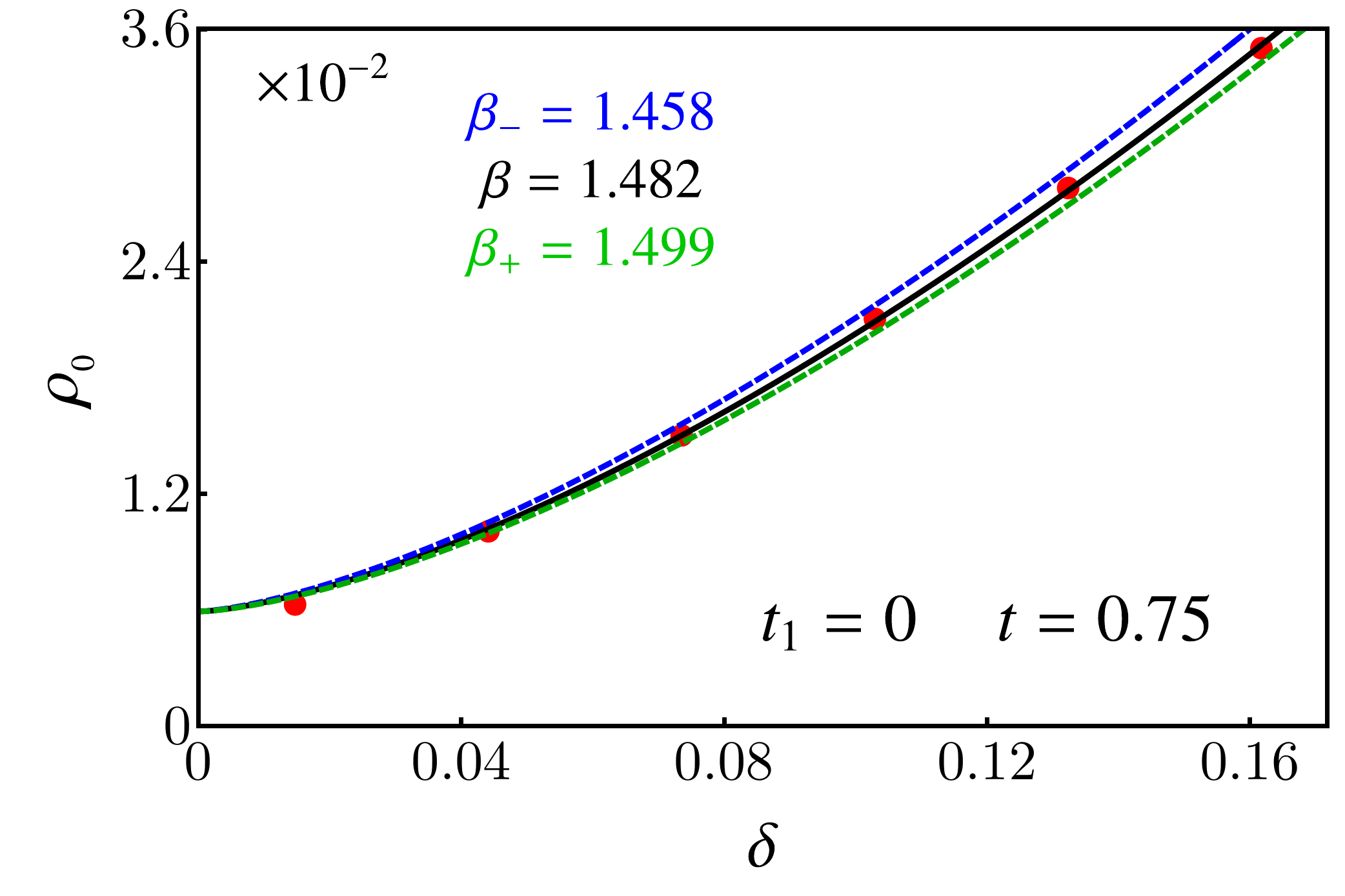}\\
\includegraphics[width=0.24\linewidth]{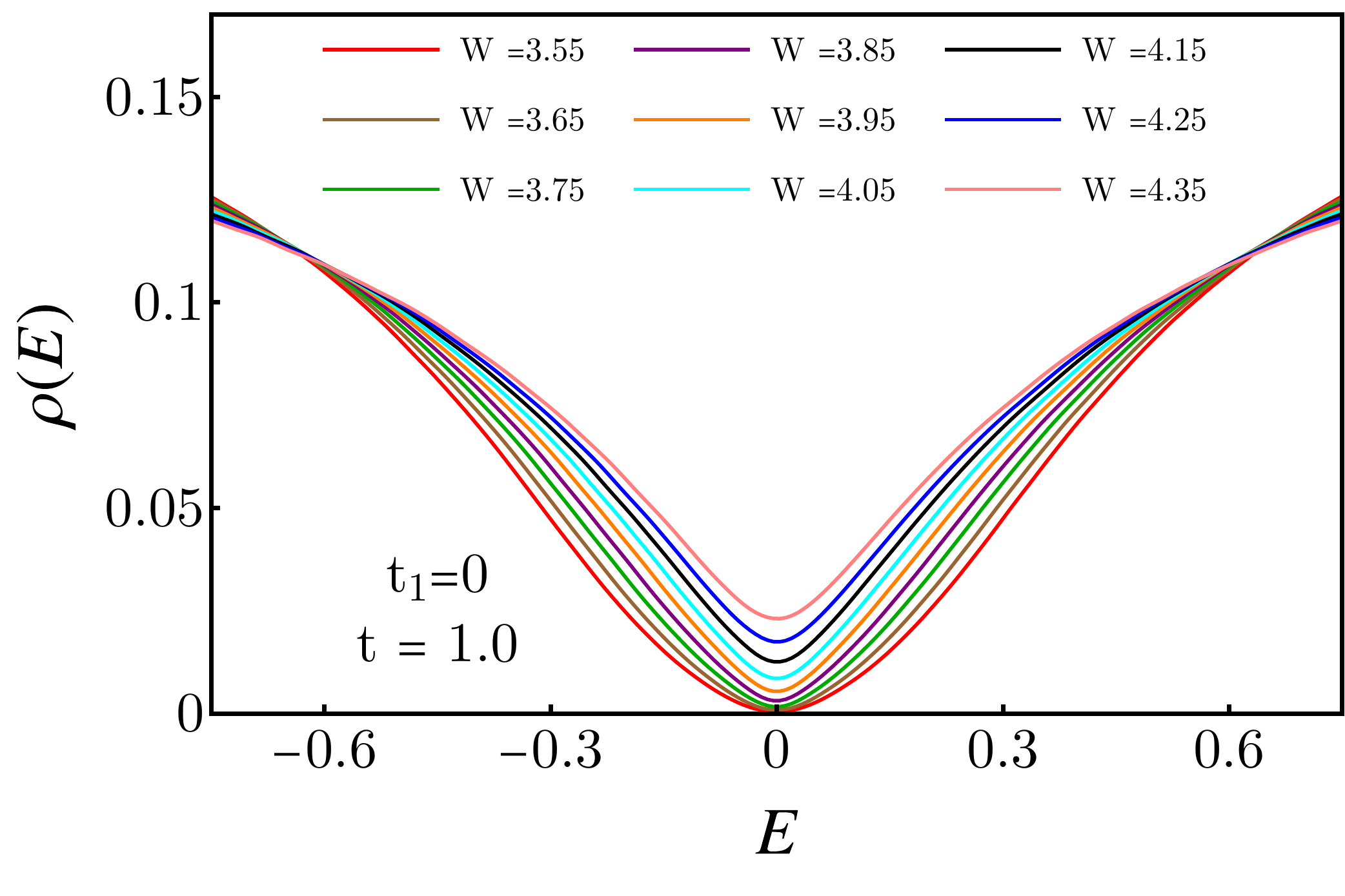}
\includegraphics[width=0.24\linewidth]{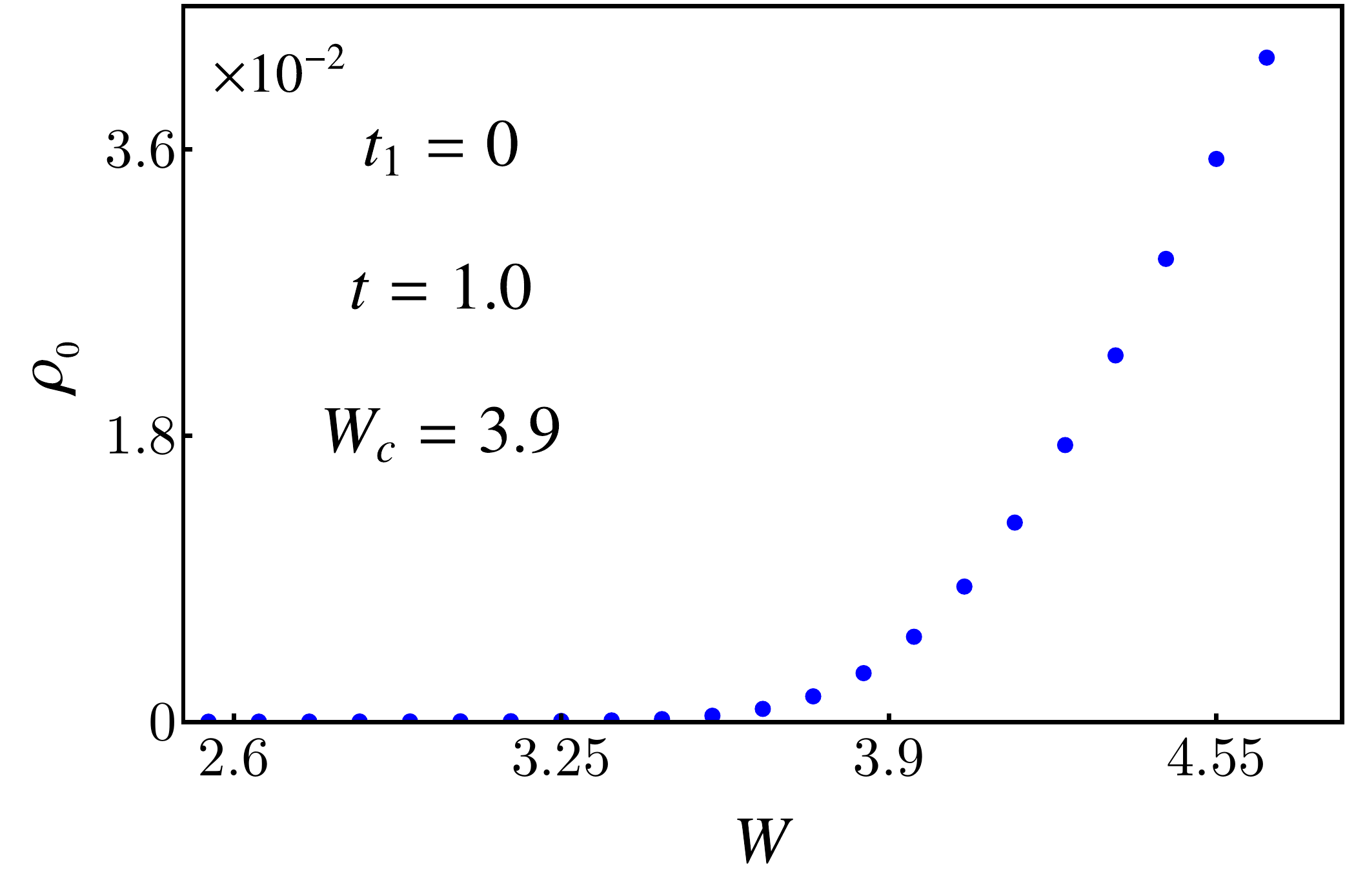}
\includegraphics[width=0.24\linewidth]{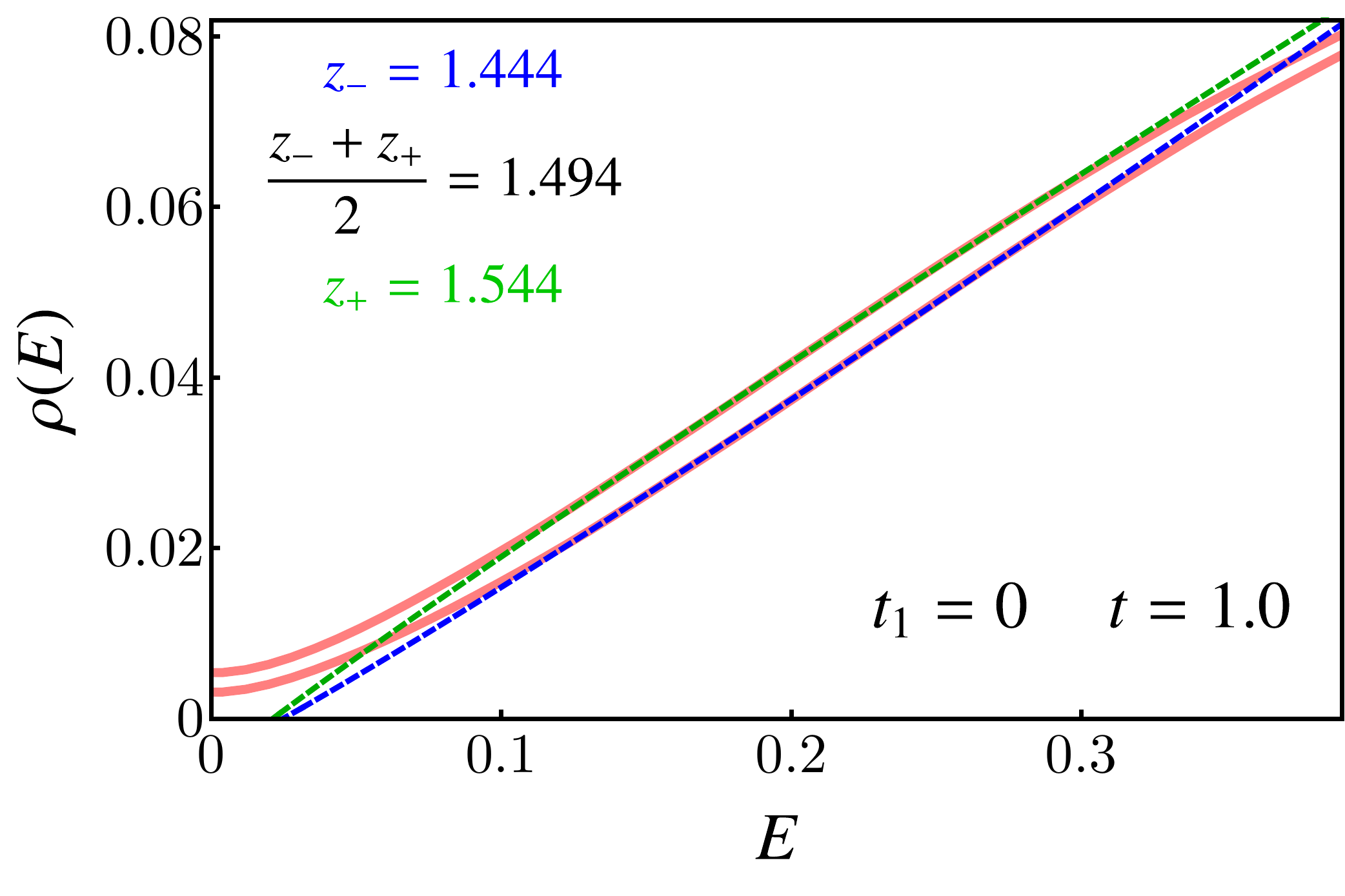}
\includegraphics[width=0.24\linewidth]{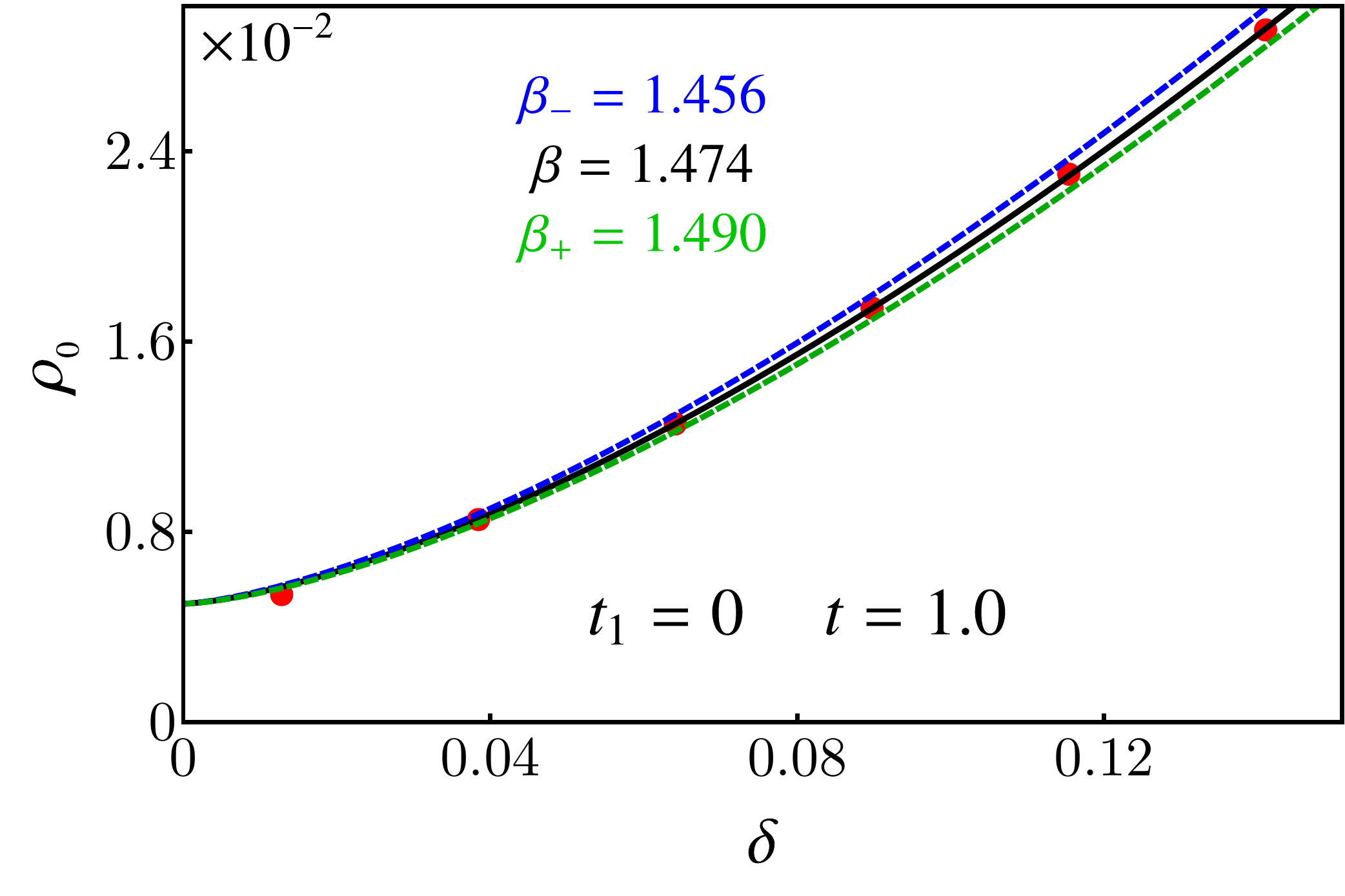}\\
\includegraphics[width=0.24\linewidth]{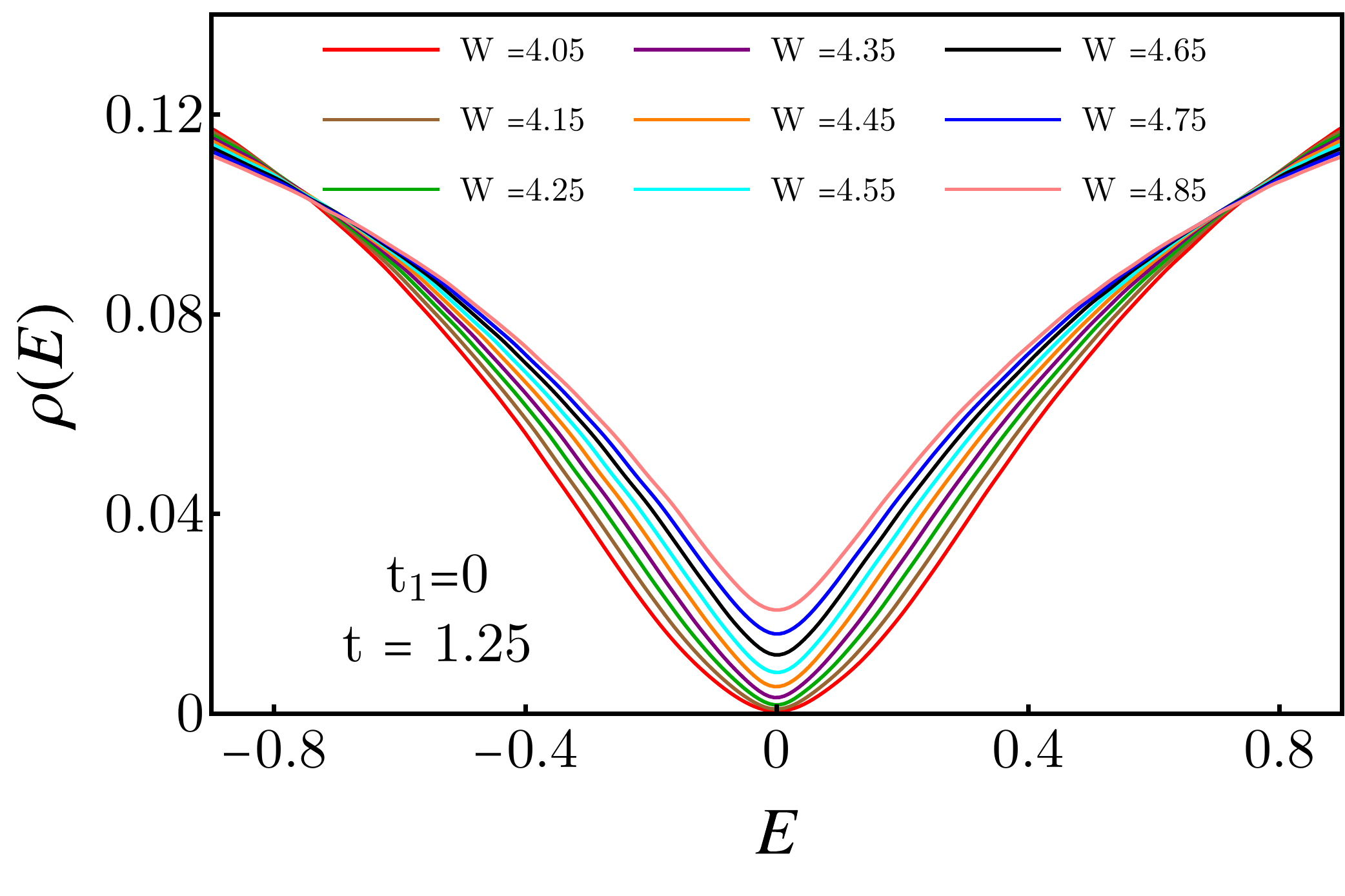}
\includegraphics[width=0.24\linewidth]{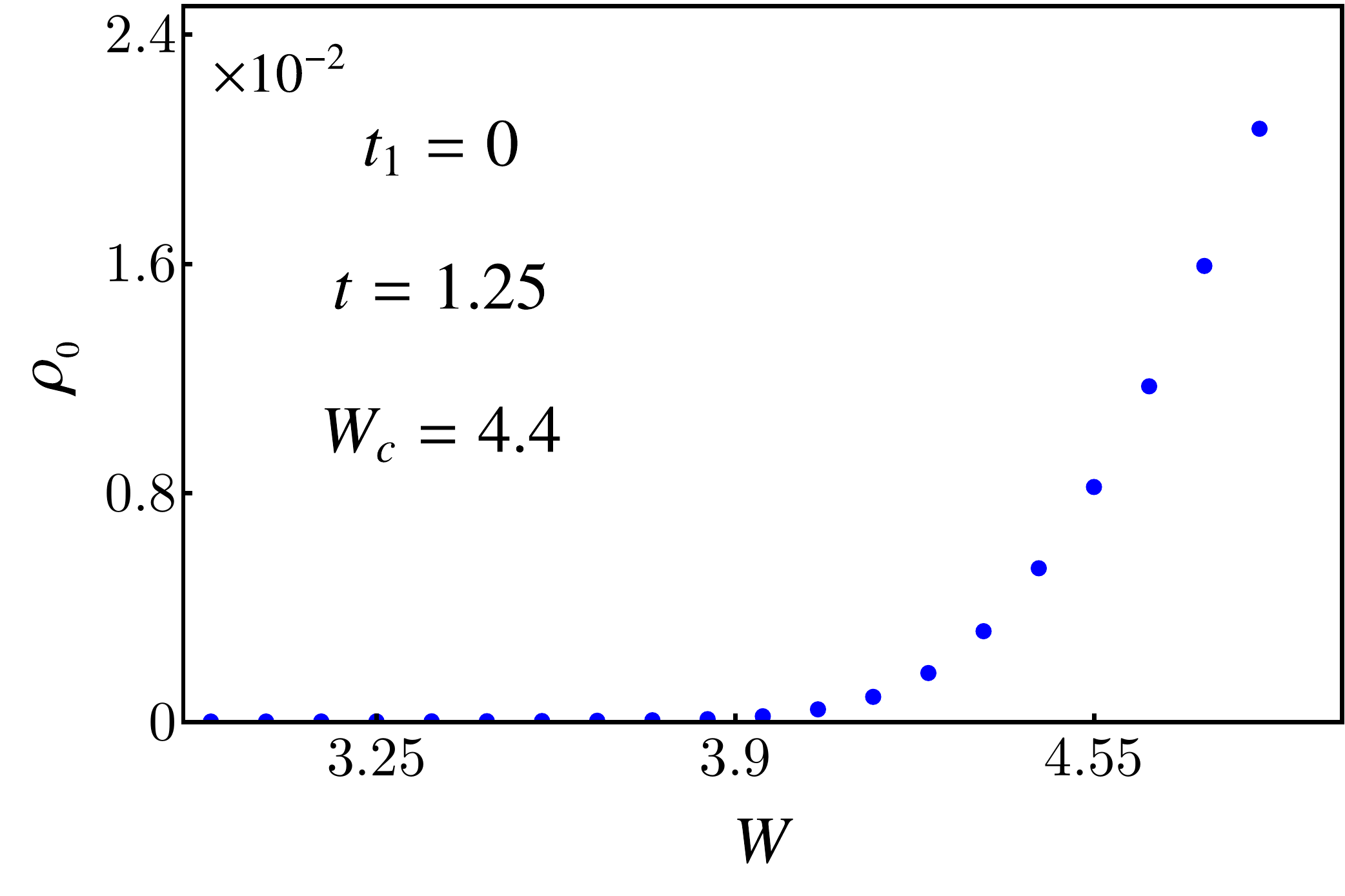}
\includegraphics[width=0.24\linewidth]{z_mInf_t05.pdf}
\includegraphics[width=0.24\linewidth]{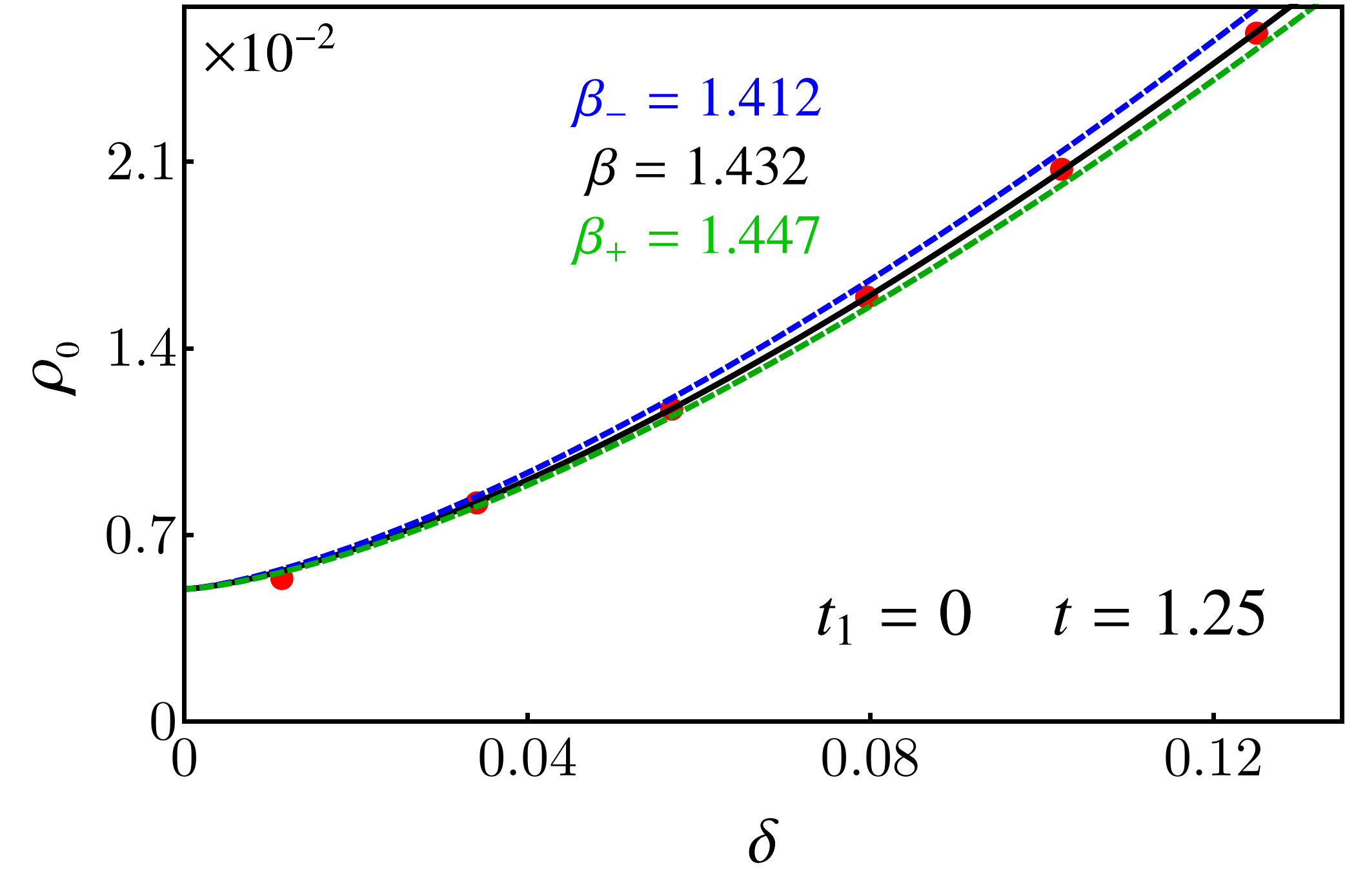}\\
\includegraphics[width=0.24\linewidth]{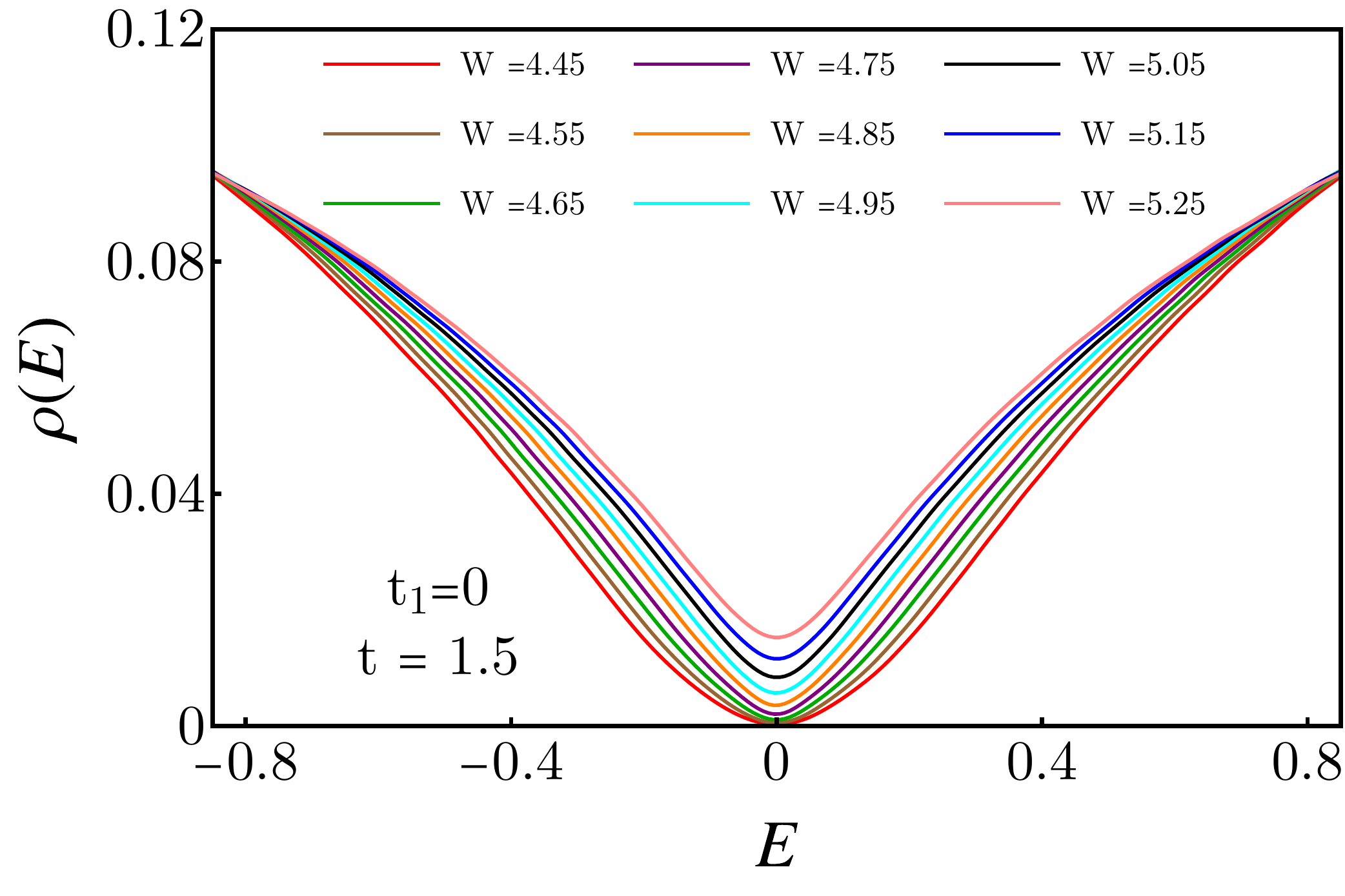}
\includegraphics[width=0.24\linewidth]{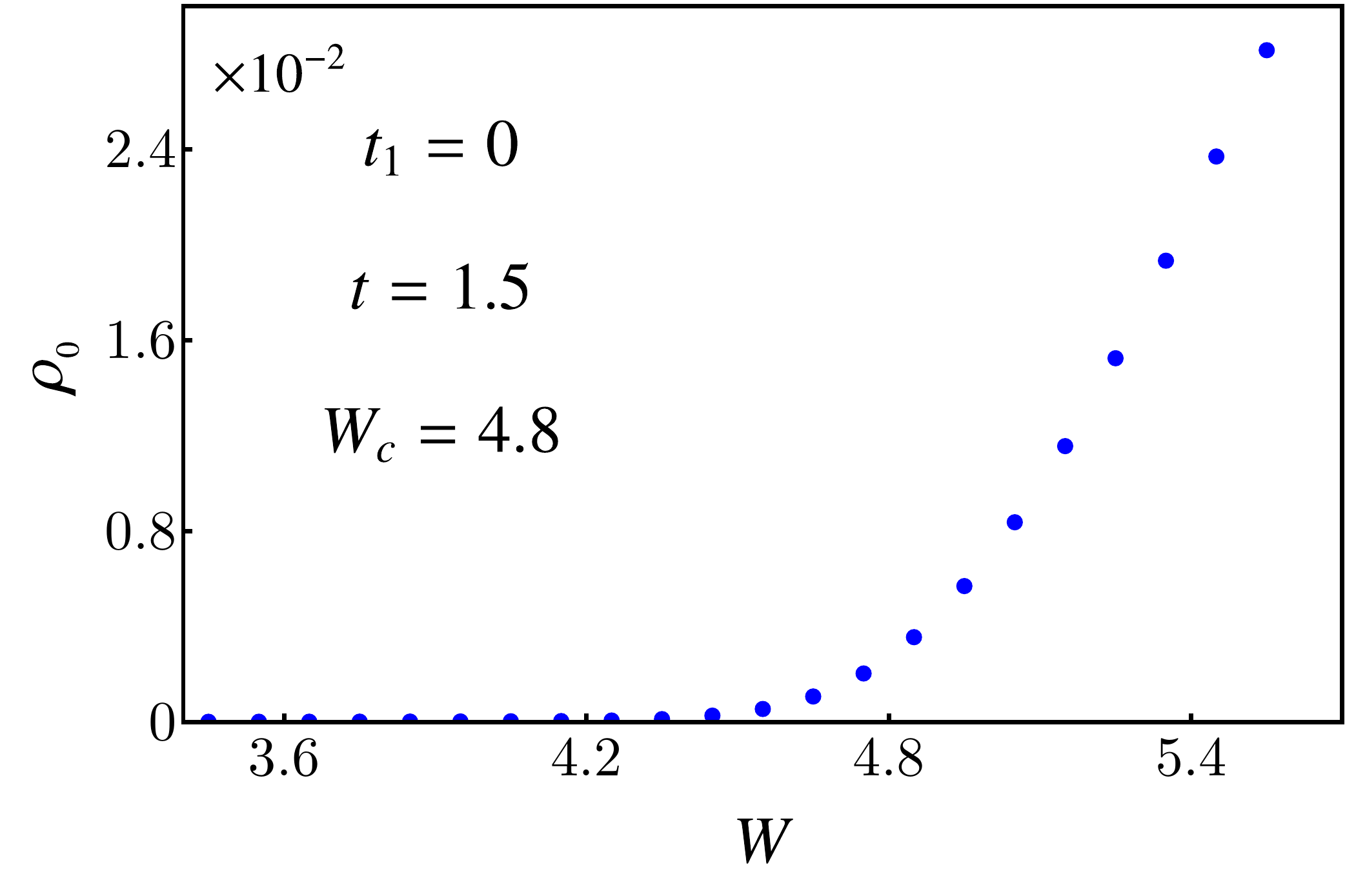}
\includegraphics[width=0.24\linewidth]{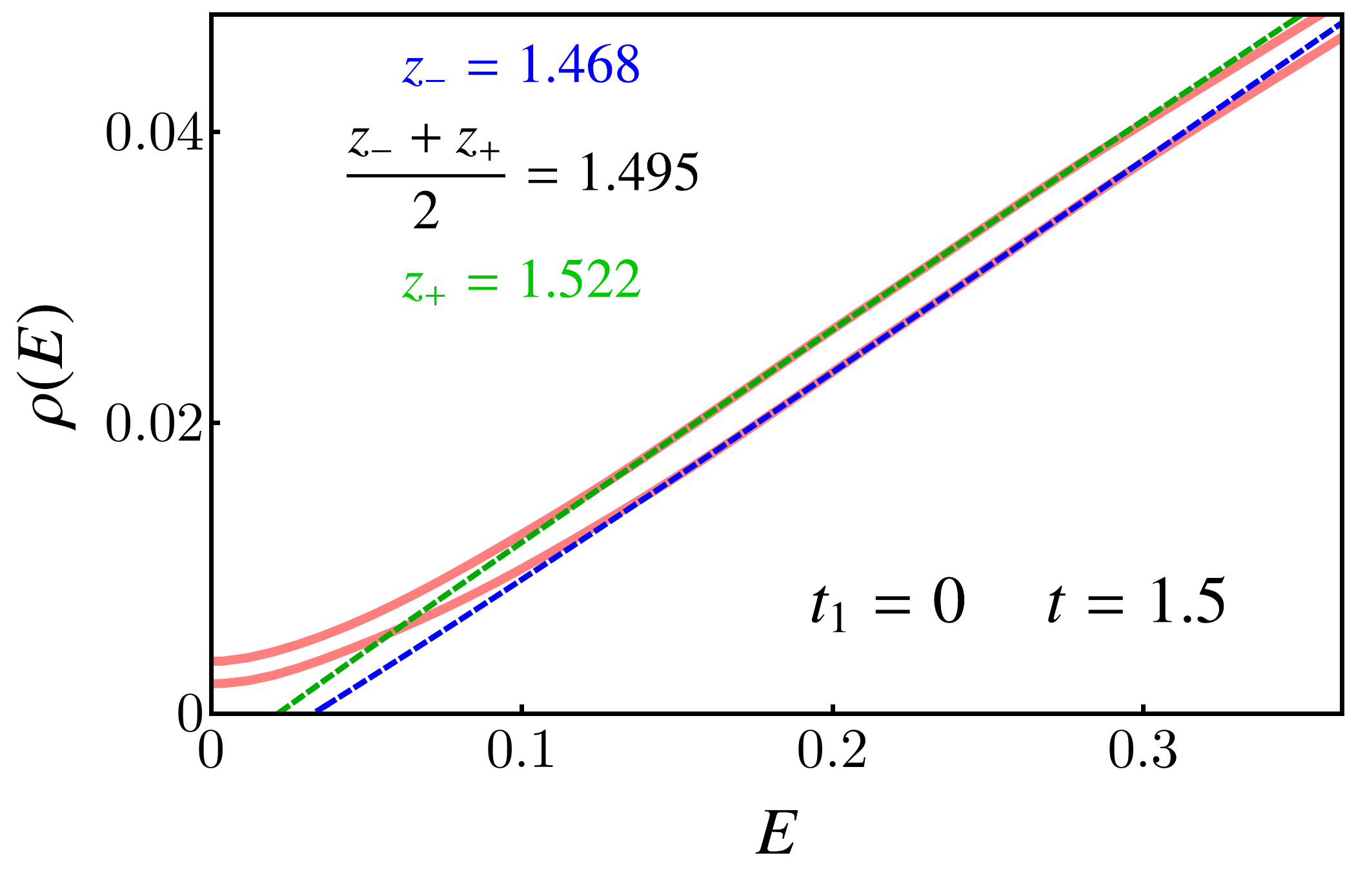}
\includegraphics[width=0.24\linewidth]{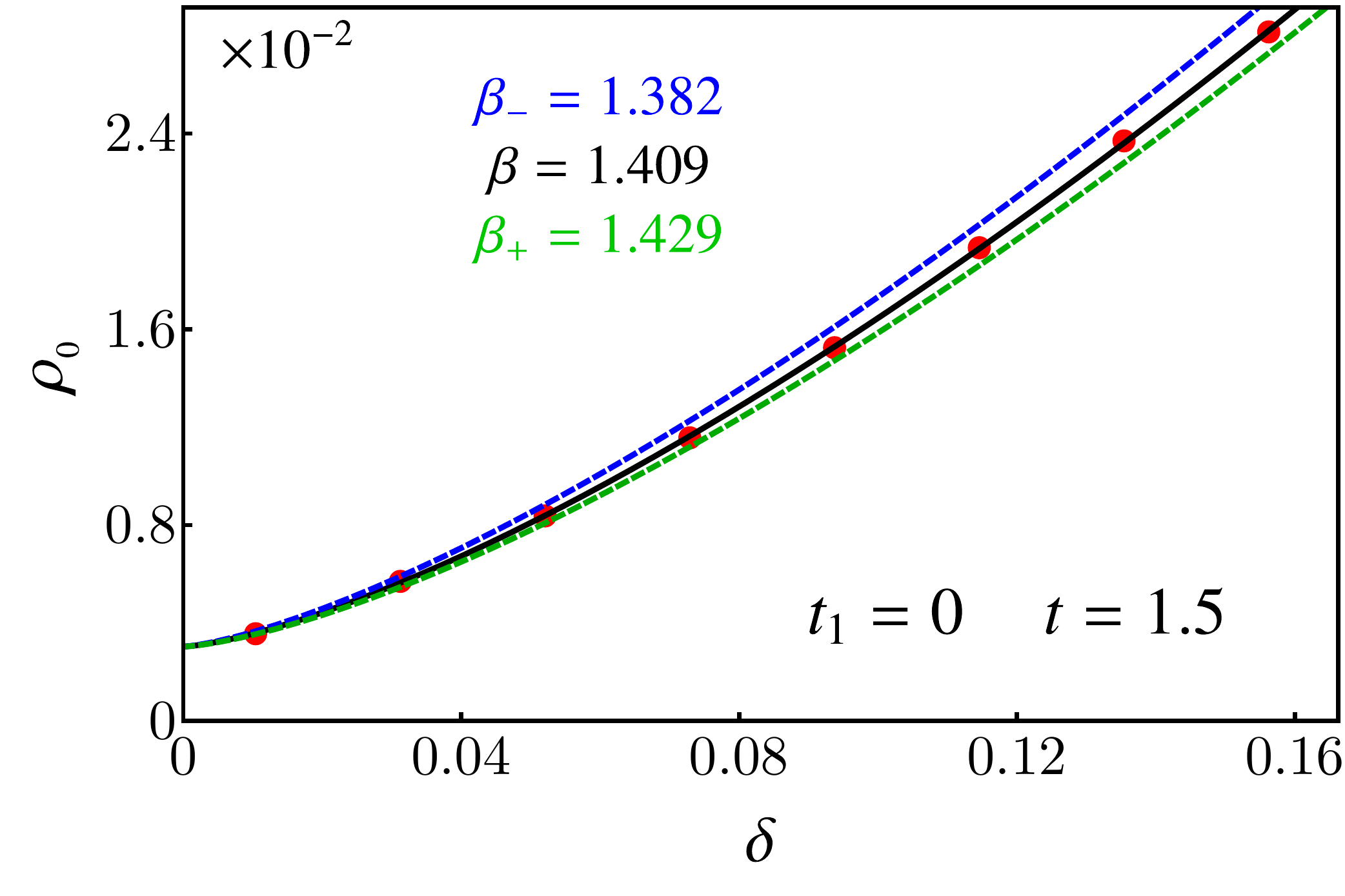}\\
\caption{Scaling analysis of DOS in a first-order DSM ($t_1=0$). With increasing $t$, the Fermi velocity increases and DOS decreases, without altering $\rho(E) \sim |E|^2$ scaling. So, $W_c$ increases with increasing $t$ (see second column). Details of the numerical analysis are discussed in Appendix~\ref{Sec:DOS_Scaling}. For additional numerical results see Ref.~\cite{SM}. 
}~\label{Fig:Dos-1}
\end{figure}

\section{Details of Numerical analysis}~\label{Sec:DOS_Scaling}

We extract the critical exponents, namely $z$ and $\nu$, across the (HOT)DSM-metal QPT by analyzing the scaling of DOS (computed using the KPM) in the following way.

{\bf Density of states}: We reconstruct the DOS from the lattice model [see Eq.~(\ref{Eq:tb})] on a cubic lattice of linear dimension $L=200$ and with periodic boundaries in each direction, see first column of Fig.~\ref{Fig:Dos-1} and Figs.~S1-S4 of the Supplementary Materials~\cite{SM}. For each $(t_1,t)$ point in parameter space, where $t_1\in \{ 0, 0.5, 1.0, 1.5, 2.0 \}$ and $t\in \{0.5,0.75,1.0,1.25,1.5\}$, we sample the disorder axis with a resolution $\Delta W=0.1$. For convenience, we choose the disorder values such that $W=n \Delta W+0.05$, where $n\in \mathbb{Z}$.

{\bf Critical disorder}: The first step in the analysis is to extract the critical disorder strength $W_c$, where the order parameter $\rho_0\equiv \rho(0)$ deviates from zero, see second column of Fig.~\ref{Fig:Dos-1} and Figs.~S1-S4 of the Supplementary Material~\cite{SM}. In practice we find the closest value that satisfies $W_c=n \Delta W$, $n\in \mathbb{Z}$. The shift between sampling $W$ and determining $W_c$ ensures (without the loss of objectivity) that no data set is exactly at $\delta=0$, which would lead to divergences in the finite energy data collapse. \\

{\bf Dynamic critical exponent}: The dynamical critical exponent $z$ is obtained by directly fitting the low-energy DOS with the function $\rho(E)\sim E^{d/z-1}$ with $d=3$. Again, in practice we fit the two adjacent data sets for $W=W_c \pm 0.05$, then take the average of the obtained values $z_-$ and $z_+$, which also serve as fitting error bars in the computation of $z$,  see third column of Fig.~\ref{Fig:Dos-1} and Figs.~S1-S4 of the Supplementary Materials~\cite{SM}.

{\bf Correlation length exponent}: As the next step we extract the order parameter critical exponent $\beta$, by fitting the function $\rho_0(\delta)\sim\delta^\beta$ for $0<\delta \ll 1$, where $\delta=(W-W_c)/W_c$ is the reduced distance from the critical point and $\beta=(d-z)\nu$. Finite size effects lead to $\rho_0(W)>0$ already for $W=W_c$, resulting in $\rho_0(\delta)$ not going to exactly zero for $\delta=0$, hence we introduce a non-zero intercept that is subject to fitting, but bounded from above by the first $\rho(\delta)$ data point, see fourth column of Fig.~\ref{Fig:Dos-1} and Figs.~S1-S4 of the Supplementary Materials~\cite{SM}. Also, we compute $\beta_\pm$ for which the resulting curves envelope all data points that were used for fitting. The correlation length exponent and its error bars are then computed using the scaling relation $\nu_{(\pm)}=\beta/(d-z_{(\pm)})$, where the final error is taken to be $\Delta \nu=\mathrm{max}(\nu_-,\nu_+)$.

The results of the numerical analysis are summarized in Table~\ref{tab:exponents}.

\twocolumngrid



\begin{thebibliography}{}

\bibitem{kane-hasan-review} M. Z. Hasan and C. L. Kane, \textit{Colloquium:} Topological insulators, Rev. Mod. Phys. {\bf 82}, 3045 (2010).

\bibitem{Qi-zhang-review} X.-L. Qi and S.-C. Zhang, Topological insulators and superconductors, Rev. Mod. Phys. {\bf 83}, 1057 (2011).

\bibitem{bernevig-hughes-book} B. A. Bernevig and T. L. Hughes, \emph{Topological Insulators and Topological Superconductors} (Princeton University Press, Princeton, New Jersey, 2013).

\bibitem{chingkai-review} C.-K. Chiu, J. C. Y. Teo, A. P. Schnyder, and S. Ryu, Classification of topological quantum matter with symmetries,
Rev. Mod. Phys. {\bf 88}, 035005 (2016).

\bibitem{armitage-review} N. P. Armitage, E. J. Mele, A. Vishwanath, Weyl and Dirac semimetals in three-dimensional solids,
Rev. Mod. Phys. {\bf 90}, 15001 (2018).


\bibitem{benalcazar2017} W. A. Benalcazar, B. A. Bernevig, and T. L. Hughes, Quantized electric multipole insulators, Science {\bf 357}, 61 (2017).

\bibitem{schindler2018} F. Schindler, Z. Wang, M. G. Vergniory, A. M. Cook, A. Murani, S. Sengupta, A. Y. Kasumov, R. Deblock, S. Jeon, I.
Drozdov, H. Bouchiat, S. Guéron, A. Yazdani, B. A. Bernevig, and T. Neupert, Higher-order topology in bismuth, Nat. Phys. {\bf 14}, 918 (2018).

\bibitem{song2017} Z. Song, Z. Fang, and C. Fang, $(D-2)$-Dimensional Edge States of Rotation Symmetry Protected Topological States, Phys. Rev. Lett. {\bf 119}, 246402 (2017).

\bibitem{benalcazar-prb2017} W. A. Benalcazar, B. A. Bernevig, and T. L. Hughes, Electric multipole moments, topological multipole moment pumping, and chiral hinge states in crystalline insulators, Phys. Rev. B {\bf 96}, 245115 (2017).

\bibitem{langbehn2017} J. Langbehn, Y. Peng, L. Trifunovic, F. von Oppen, and P. W. Brouwer, Reflection-Symmetric Second-Order Topological Insulators and Superconductors, Phys. Rev. Lett. {\bf 119}, 246401 (2017).

\bibitem{franca2018} S. Franca, J. van den Brink, and I. C. Fulga, An anomalous higher-order topological insulator,
Phys. Rev. B {\bf 98}, 201114(R) (2018).

\bibitem{schindler-sciadv2018} F. Schindler, A. M. Cook, M. G. Vergniory, Z. Wang, S. S. P. Parkin, B. A. Bernevig, and T. Neupert, Higher-order topological insulators, Sci. Adv. {\bf 4}, eaat0346 (2018).

\bibitem{wang-PRL2019} Z. Wang, B. J. Wieder, J. Li, B. Yan, and B. A. Bernevig, Higher-Order Topology, Monopole Nodal Lines, and the Origin of Large Fermi Arcs in Transition Metal Dichalcogenides \textbf{\textit{X}Te}$_2$ (\textbf{\textit{X}=Mo,W}), Phys. Rev. Lett. {\bf 123}, 186401 (2019).

\bibitem{hsu2018} C.-H. Hsu, P. Stano, J. Klinovaja, and D. Loss, Majorana Kramers Pairs in Higher-Order Topological Insulators,
Phys. Rev. Lett. {\bf 121}, 196801 (2018).

\bibitem{wang1-2018} Y. Wang, M. Lin, and T. L. Hughes, Weak-pairing higher order topological superconductors, Phys. Rev. B {\bf 98}, 165144 (2018).

\bibitem{hughes-HOTDSM} M. Lin and T. L. Hughes, Topological quadrupolar semimetals, Phys. Rev. B {\bf 98}, 241103(R) (2018).

\bibitem{calugaru2019} D. C\u{a}lug\u{a}ru, V. Juri\v ci\' c, and B. Roy, Higher-order topological phases: A general principle of construction, Phys. Rev. B {\bf 99}, 041301(R) (2019).

\bibitem{ghorashi2019} S. A. A. Ghorashi, X. Hu, T. L. Hughes, E. Rossi, Second-order Dirac superconductors and magnetic field induced Majorana hinge modes, Phys. Rev. B {\bf 100}, 020509 (2019). 

\bibitem{Klinovaja2019} Y. Volpez, D. Loss, and J. Klinovaja, Second-Order Topological Superconductivity in $\pi$-Junction Rashba Layers,
Phys. Rev. Lett. {\bf 122}, 126402 (2019).

\bibitem{Tnag2019} T. Nag, V. Juri\v ci\' c, B. Roy, Out of equilibrium higher-order topological insulator: Floquet engineering and quench dynamics, Phys. Rev. Research {\bf 1}, 032045(R) (2019).

\bibitem{Klinovaja2019PRR} K. Plekhanov, M. Thakurathi, D. Loss, J. Klinovaja, \emph{Floquet second-order topological superconductor driven via ferromagnetic resonance}, Phys. Rev. Research {\bf 1}, 032013 (2019).

\bibitem{ZYan2019} Z. Yan, Higher-Order Topological Odd-Parity Superconductors, Phys. Rev. Lett. {\bf 123}, 177001 (2019).

\bibitem{roy-singleauthor2019} B. Roy, Antiunitary symmetry protected higher-order topological phases, Phys. Rev. Research {\bf 1}, 032048(R) (2019).

\bibitem{bernevig-NatComm2020} B. J. Wieder, Z. Wang, J. Cano, X. Dai, L. M. Schoop, B. Bradlyn, B. A. Bernevig, Strong and fragile topological Dirac semimetals with higher-order Fermi arcs, Nat. Commun. {\bf 11}, 627 (2020).

\bibitem{VLiu-PRL-2020} H. Hu, B. Huang, E. Zhao, W. V. Liu, Dynamical Singularities of Floquet Higher-Order Topological Insulators,
Phys. Rev. Lett. {\bf 124}, 057001 (2020).

\bibitem{andras-2019} A. Szab\' o, R. Moessner, and B. Roy, Strain-engineered higher-order topological phases for spin-3/2 Luttinger fermions, Phys. Rev. B {\bf 101}, 121301(R) (2020).

\bibitem{dassarma-HOTSC} R.-X. Zhang, Y.-T. Hsu, and S. Das Sarma, Higher-order topological Dirac superconductors,
Phys. Rev. B {\bf 102}, 094503 (2020).



\bibitem{KPM-RMP} A. Wei\ss{}e, G. Wellein, A. Alverman, and H. Feshke, The kernel polynomial method, Rev. Mod. Phys. {\bf 78}, 275 (2006).

\bibitem{SM} See the Supplemental Materials at XXX-XXXX for additional numerical results. 

\bibitem{jackiw-rebbi} R. Jackiw and C. Rebbi, Solitons with fermion number 1/2, Phys. Rev. D {\bf 13}, 3398 (1976).


\bibitem{fradkin} E. Fradkin, Critical behavior of disordered degenerate semiconductors. II. Spectrum and transport properties in mean-field theory, Phys. Rev. B {\bf 33}, 3263 (1986).

\bibitem{goswami-chakravarty-1} P. Goswami and S. Chakravarty, Quantum Criticality between Topological and Band Insulators in 3+1 Dimensions, 
Phys. Rev. Lett. {\bf 107}, 196803 (2011).

\bibitem{ominato-kishino} Y. Ominato and M. Koshino, Quantum transport in a three-dimensional Weyl electron system, Phys. Rev. B {\bf 89}, 054202 (2014).

\bibitem{herbut} K. Kobayashi, T. Ohtsuki, K.-I. Imura, and I. F. Herbut, Density of States Scaling at the Semimetal to Metal Transition in Three Dimensional Topological Insulators, Phys. Rev. Lett. {\bf 112}, 016402 (2014).

\bibitem{roy-dassarma-1} B. Roy and S. Das Sarma, Diffusive quantum criticality in three-dimensional disordered Dirac semimetals, 
Phys. Rev. B {\bf 90}, 241112(R) (2014); \emph{ibid}. {\bf 93}, 119911(E) (2016).

\bibitem{brouwer-1} B. Sbierski, G. Pohl, E. J. Bergholtz, and P. W. Brouwer, Quantum Transport of Disordered Weyl Semimetals at the Nodal Point, Phys. Rev. Lett. {\bf 113}, 026602 (2014).

\bibitem{syzranov-1} S. V. Syzranov, L. Radzihovsky, and V. Gurarie, Critical Transport in Weakly Disordered Semiconductors and Semimetals, 
Phys. Rev. Lett. {\bf 114}, 166601 (2015).

\bibitem{brouwer-2} B. Sbierski, E. J. Bergholtz, and P. W. Brouwer, Quantum critical exponents for a disordered three-dimensional Weyl node, Phys. Rev. B {\bf 92}, 115145 (2015).

\bibitem{pixley-1} J. H. Pixley, P. Goswami, and S. Das Sarma, Anderson Localization and the Quantum Phase Diagram of Three Dimensional Disordered Dirac Semimetals, Phys. Rev.Lett. {\bf 115}, 076601 (2015).

\bibitem{syzranov-2}  S. V. Syzranov, P. M. Ostrovsky, V. Gurarie, and L. Radzihovsky, Critical exponents at the unconventional disorder-driven transition in a Weyl semimetal, Phys. Rev. B {\bf 93}, 155113 (2016). 

\bibitem{pixley-2} J. H. Pixley, P. Goswami, and S. Das Sarma, Disorder-driven itinerant quantum criticality of three-dimensional massless Dirac fermions, Phys. Rev. B {\bf 93}, 085103 (2016).

\bibitem{ohtsuki-shindou} S. Liu, T. Ohtsuki, and R. Shindou, Effect of Disorder in a Three-Dimensional Layered Chern Insulator, 
Phys. Rev. Lett. {\bf 116}, 066401 (2016).

\bibitem{bera-sau-roy} S. Bera, J. D. Sau, and B. Roy, Dirty Weyl semimetals: Stability, phase transition, and quantum criticality, 
Phys. Rev. B {\bf 93}, 201302(R) (2016).

\bibitem{roy-dassarma-2} B. Roy, V. Juri\v ci\' c, and S. Das Sarma, Universal optical conductivity of a disordered Weyl semimetal, 
Sci. Rep. {\bf 6}, 32446 (2016).

\bibitem{carpentier-1} T. Louvet, D. Carpentier, and A. A. Fedorenko, On the disorder-driven quantum transition in three-dimensional relativistic metals, Phys. Rev. B {\bf 94}, 220201(R) (2016).

\bibitem{goswami-chakravarty-2} P. Goswami and S. Chakravarty, Superuniversality of topological quantum phase transition and global phase diagram of dirty topological systems in three dimensions, Phys. Rev. B {\bf 95}, 075131 (2017).

\bibitem{carpentier-2} T. Louvet, D. Carpentier, and A. A. Fedorenko, New quantum transition in Weyl semimetals with correlated disorder, 
Phys. Rev. B {\bf 95}, 014204 (2017).

\bibitem{larsfritz} T. S. Sikkenk and L. Fritz, Disorder in tilted Weyl semimetals from a renormalization group perspective, Phys. Rev. B {\bf 96}, 155121 (2017).

\bibitem{roy-slager-juricic-1} B. Roy, R-J. Slager, and V. Juri\v ci\' c, Global Phase Diagram of a Dirty Weyl Liquid and Emergent Superuniversality, Phys. Rev. X {\bf 8}, 031076 (2018). 

\bibitem{carpetier-3} I. Balog, D. Carpentier, and A. A. Fedorenko, Disorder-Driven Quantum Transition in Relativistic Semimetals: Functional Renormalization via the Porous Medium Equation, Phys. Rev. Lett. {\bf 121}, 166402 (2018).

\bibitem{mirlin} J. Klier, I. V. Gornyi, and A. D. Mirlin, From weak to strong disorder in Weyl semimetals: Self-consistent Born approximation, Phys. Rev. B {\bf 100}, 125160 (2019). 

\bibitem{carpentier-4} E. Brillaux, D. Carpentier, and A. A. Fedorenko, Multifractality at the Weyl-semimetal–diffusive-metal transition for generic disorder, Phys. Rev. B {\bf 100}, 134204 (2019). 

\bibitem{ogata} T. Hirosawa, H. Maebashi, M. Ogata, Nuclear spin relaxation rate near the disorder-driven quantum critical point in Weyl fermion systems, Phys. Rev. B {\bf 101}, 155103 (2020).


\bibitem{nandkishore-RR}  R. Nandkishore, D. A. Huse, and S. L. Sondhi, Rare region effects dominate weakly disordered three-dimensional Dirac points, Phys. Rev. B {\bf 89}, 245110 (2014).

\bibitem{huses-pixley-RR} J. H. Pixley, D. A. Huse, and S. Das Sarma, Rare-Region-Induced Avoided Quantum Criticality in Disordered Three-Dimensional Dirac and Weyl Semimetals, Phys. Rev. X {\bf 6}, 021042 (2016).

\bibitem{altland-2} M. Buchhold, S. Diehl, and A. Altland, Vanishing Density of States in Weakly Disordered Weyl Semimetals, 
Phys. Rev. Lett. {\bf 121}, 215301 (2018). 



\bibitem{roy-slager-juricic-2} R.-J. Slager, V. Juri\v ci\' c, and B. Roy, Dissolution of topological Fermi arcs in a dirty Weyl semimetal, 
Phys.Rev. B {\bf 96}, 201401 (2017).


\end{thebibliography}
\end{document}